\documentclass[twocolumn]{aastex631}
\usepackage{amsmath}
\usepackage{wasysym}
\usepackage{graphicx}
\usepackage{amssymb}
\usepackage{epstopdf}
\usepackage{mathrsfs}
\usepackage{anyfontsize}
\usepackage{natbib}
\usepackage{color}
\usepackage{lipsum}
\usepackage{diagbox}
\usepackage{tabularx}
\usepackage{booktabs}
\usepackage{makecell}
\usepackage{hyperref}

\newcommand{\Msun}{M_{\odot}}
\newcommand{\etash}{\eta_{\rm sh}}
\defcitealias{coughlin23}{C23}
\defcitealias{Fernandez18}{F18}

\shorttitle{Neutrino Mass Loss Shocks}
\shortauthors{Paradiso, Vallejo, \& Coughlin}
\begin{document}   
\title{On the Origin of Mass Ejection in Failed Supernovae}
\author[0009-0003-8285-0702]{Daniel A.~Paradiso}
\affiliation{Department of Physics, Syracuse University, Syracuse, NY 13210, USA}
\author[0009-0006-8046-4156]{Sarah Vallejo}
\affiliation{Department of Physics, Syracuse University, Syracuse, NY 13210, USA}
\author[0000-0003-3765-6401]{Eric R.~Coughlin}
\affiliation{Department of Physics, Syracuse University, Syracuse, NY 13210, USA}

\email{dparadis@syr.edu}
\email{ecoughli@syr.edu}

\begin{abstract}
Some high-mass stars likely end their lives in underluminous implosions that leave behind a black hole, known as failed supernovae (FSNe). However, neutrinos radiated during proto-neutron star formation generate a weak (Mach $\gtrsim 1$) shockwave in the outer layers of the star, which produces a unique transient as it breaks out of the dying star and signals its imminent disappearance. It was recently shown that there are two self-similar solutions that describe the propagation of this weak shockwave, and these solutions simultaneously contain outward-moving ejecta and fallback accretion onto the black hole. Here we show that the larger Mach number solutions are unstable, such that the Mach number of the shock grows with time $t$ and deviates from the self-similar prediction as $\propto t^{\alpha}$, with $\alpha \lesssim 0.1$, whereas the smaller Mach number solutions are stable. We also show that, above a critical mass loss that is readily achievable in core-collapse supernovae, the shock asymptotically strengthens and approaches the strong limit. Our results imply that it is the mass lost to neutrinos \textit{relative} to the mass enclosed by the shockwave, as well as the stellar density gradient where the shock forms, that primarily dictate its strength and the amount of material it ejects. These criteria explain why red supergiants, which have relative mass losses well in excess of the critical value at the time of shock formation, more readily eject material and create more luminous explosions compared to more compact progenitors. 
\end{abstract}


\section{Introduction}
\label{sec: intro}
The birth of stellar mass black holes from failed supernovae (FSNe) -- the unsuccessful explosions resulting from the core collapse of massive ($\gtrsim 8 \Msun$) stars -- is indirectly evidenced by, for example, discrepancies between high mass star formation and core-collapse supernova (CCSN) rates \citep{horiuchi11, smartt15}, the observed mass distribution of compact objects (e.g., \citealt{Kochanek14,Kochanek15, Raithel18, Disberg23, LIGO25, Willcox25, Legred26}), and numerical simulations of CCSNe (e.g., \citealt{blondin03, sukhbold16, Muller17, Burrows24}), and represent plausible progenitors of some luminous fast blue optical transients (LFBOTs; e.g., \citealt{Drout14, Kashiyama15, Margutti19, Coppejans20, Ho20, Yao20, Chrimes26, LeBaron26, Perley26}). Two supergiant stars in nearby galaxies have also recently been observed fading in luminosity before disappearing without an associated CCSN, tentatively making them the first direct detections of FSNe and stellar mass black hole formation (\citealt{Gerke15, Adams17, Basinger21, Kochanek24,De26, De26b, ForesToribio26}; but see \citealt{Beasor24, Beasor26, Soker26}). In particular, observations and theory suggest that it is high mass ($\gtrsim 18 \Msun$; \citealt{smartt15}) stars with a high central mass concentration --- commonly referred to as ``compactness'' --- at the onset of core-collapse that are prone to failure \citep{OConnor11, sukhbold16, Ertl16, Maltsev25}. Therefore, unless the rotation of the progenitor facilitates the formation of a disk --- which could lead to the launching of jets in an accretion-powered transient (e.g., \citealt{Woosley93, MacFadyen99, Quataert12, Gilkis16b}) --- these compact stars experience a higher accretion rate onto the neutron star formed during core-collapse, thereby expediting black hole formation and precluding traditional stellar explosion.

Nevertheless, even these ``failed'' explosions are capable of producing a transient that signifies the death of the star and formation of a solar mass black hole, i.e., failed supernovae are not identically ``disappearing stars.'' Specifically, neutrino emission during neutron star formation reduces the gravitational mass of the core by $\sim 0.1-0.5\Msun$ \citep{Burrows88, Lattimer89}. While $\sim 1\%$ of these weakly interacting neutrinos deposit energy in the region behind the stalled neutron star bounce shock --- i.e., the typical ``neutrino explosion mechanism'' (e.g., \citealt{Colgate66, Bethe85}) --- the vast majority stream unimpeded through the overlying stellar envelope. In the outer regions of the star, this reduction in the gravitational potential of the core occurs effectively instantaneously, as the time it takes for neutrinos to stream out of the progenitor --- the combination of the neutrino diffusion time ($\sim1-10~\rm seconds$; e.g., \citealt{Burrows86, Pons99}) and the light crossing time ($\sim$ hours) --- is negligible in comparison to the local dynamical time near the stellar surface ($\sim$ weeks to months). At these large radii the stellar envelope then responds dynamically and outwardly to this liberation of mass-energy, and this outward-moving region is connected to the inner and infalling region (where material is accreting onto the black hole) via a second, relatively weak (Mach number order unity) shockwave (e.g., \citealt{Lovegrove13, Coughlin18b, Fernandez18}), a mechanism first considered by \cite{Nadezhin80}.

The formation of these weak, neutrino mass loss-induced shockwaves and their potential to unbind the envelopes of supergiant stars has since been studied analytically (e.g., \citealt{Piro13,Coughlin18b, coughlin18, coughlin19}) and numerically (e.g., \citealt{Lovegrove13, Lovegrove17, Fernandez18, Tsuna20, Ivanov21, Schneider23, Antoni25}). This work has shown that the result of the explosion is heavily dependent on the type of supergiant, the overall stellar structure, and the neutron star equation of state \citep{Fernandez18, Ivanov21, Schneider23}. Specifically, stars with extended and shallow envelopes, as is common in red supergiants, are conducive to generating longer-duration explosions with up to a few solar masses of material ejected, whereas more compact stars such as blue and yellow supergiants generate shorter duration transients with only a $few\times0.1\Msun$ of ejecta \citep{Fernandez18}. Additionally, it has been shown that angular momentum due to convective motions in the outer regions of the infalling stellar envelope can inhibit accretion onto the nascent black hole, leading to the formation of a disk, launching of jets, and unbinding of the stellar envelope (e.g., \citealt{Gilkis16, Quataert19, Antoni22, Antoni23}).

The propagation of a weak shockwave in a point-mass gravitational field of mass $M$ that forms in response to an instantaneous loss of mass $\delta M$ was shown to have a self-similar solution by Coughlin (\citeyear{coughlin23}; hereafter \citetalias{coughlin23}), provided that the relative neutrino mass loss $\delta M/M$ is below a critical value that is typically a few percent the mass of the star. These shockwaves have Mach numbers of at most a few, exhibit accretion onto the nascent black hole, and are dependent on the relative amount of mass lost to neutrinos, $\delta M/M$, for which two self-similar solutions exist below the critical value. We characterize these solutions by their Mach number, and refer to them as the ``strong'' and ``weak'' solutions, with the strong (weak) solution having the larger (smaller) Mach number of the two. These solutions also depend on the ambient medium into which the shock is propagating, which is non-hydrostatic as a consequence of the neutrino mass loss, with the self-similar ambient fluid profiles being characterized by the ambient density power-law index $n$, adiabatic index $\gamma$, and relative mass loss $\delta M/M$. 

As was discussed in \citetalias{coughlin23}, the strong shock solutions are generalizations of the self-similar solutions found by \citet{coughlin18} in the limit that $\delta M / M\rightarrow0$, whereas the weak solutions converge to the rarefaction wave self-similar solution of \citealt{coughlin19} (or the solution of \citealt{Kazhdan94} for $n\leq2$) in the same zero mass loss limit. \citealt{coughlin23} hypothesized that the strong shock solutions are unstable to radial perturbations (i.e., perturbations to the self-similar solution grow with time, as was shown to be the case for the \citealt{coughlin18} solutions in \citealt{coughlin19}), while the weak shock solutions are stable since \citealt{coughlin19} showed that the rarefaction wave solutions are stable. As neither solution exists when the mass loss is above the critical value, \citetalias{coughlin23} conjectured that, in this regime of ``super-critical mass loss,'' the shockwave is effectively unstable from the standpoint that it will asymptotically strengthen over time and transition into the Sedov-Taylor/energy-conserving blastwave \citep{Taylor50, Sedov59}. 

Here we develop a formalism to rigorously assess the stability of the solutions found in \citetalias{coughlin23} in both the sub- and super-critical mass loss regimes. We begin with the sub-critical mass loss solutions in Section \ref{sec: local} and show that the weak and strong shock solutions are indeed stable and unstable, respectively, and we confirm these results with numerical simulations using the hydrodynamics code \textsc{flash} \citep{fryxell00}. We additionally extend the formalism of Section \ref{sec: local} to incorporate non-radial perturbations in Appendix \ref{sec: Angular}, which we show are stable for spherical harmonic $\ell>0$ for both the strong and weak shock solutions. In Section \ref{sec: global} we show that solutions with mass losses above the critical value can be modeled by perturbing the maximum mass loss self-similar solution, with perturbations being driven by the additional mass lost above the critical value. Unlike the sub-critical mass loss solutions, which have perturbations that grow as weak power-laws in time, these perturbations grow logarithmically with the shock position and the shock strengthens with time. 

In Section \ref{sec: implications} we discuss the implications of our findings in the context of realistic stellar progenitors, and -- based on the results of Sections \ref{sec: local} \& \ref{sec: global} -- we suggest that it is largely the \emph{relative} mass loss and the stellar density gradient at the radius where the shockwave forms that influence its propagation. As such, less compact stars like red supergiants -- which have large relative mass losses in their interiors -- are expected to produce strong explosions and substantial ($\gtrsim 1 M_{\odot}$) ejecta compared to more compact blue supergiants, yellow supergiants, and Wolf-Rayets, and these outcomes agree with the results of prior numerical investigations (e.g., \citealt{Lovegrove13, Lovegrove17, Fernandez18, Ivanov21, Schneider23}). We summarize and conclude in Section \ref{sec: sum}. 

\section{Local Stability of Self-similar Solutions}
\label{sec: local}
\begin{table*}
\centering
\caption{Variables Defined in Paper I. Note that the ambient solutions are defined in terms of the Eulerian radius $r$, whereas the post-shock solutions are written relative to the shock position and velocity.}
\label{tab:1}
\begin{tabular}{lcc}
\toprule
Variable & \multicolumn{1}{c}{Ambient} & \multicolumn{1}{c}{Post-Shock} \\
\midrule
Neutrino mass loss & $\delta M$ & $\delta M$ \\
Point mass & $M$ & $M$ \\
Ambient length scale, density & $r_{\rm i}, \rho_{\rm i}$ & $r_{\rm i}, \rho_{\rm i}$ \\
Time-dependent shock position, velocity & $---$ & $R_{\rm sh}, V_{\rm sh}$ \\
Self-Similar variable & $\eta = \sqrt{GMt}/r^{3/2}$ & $\xi = r/R_{\rm sh}$ \\
Implicit shock position & $---$ & $\eta_{\rm sh} = \sqrt{GMt}/R_{\rm sh}^{3/2}$ \\
Self-similar shock velocity & $---$ & $\left(2/3\right){\eta_{\rm sh}}^{-1}\sqrt{GM/R_{\rm sh}}$ \\
Self-similar fluid velocity & $v_e = \sqrt{GM/r} f_e\left(\eta\right)$ & $v = V_{\rm sh}\left(t\right) f\left(\xi\right)$ \\
Self-similar fluid density & $\rho_e = \rho_i (r/r_{\rm i})^{-n} g_e\left(\eta\right)$ & $\rho = \rho_{\rm i} (R_{\rm sh}\left(t\right)/r_{\rm i})^{-n} g\left(\xi\right)$ \\
Self-similar fluid pressure & $p_e = (n+1)^{-1} \rho_{\rm i} (GM/r) (r/r_{\rm i})^{-n} h_e\left(\eta\right)$ & 
$p = \rho_{\rm i} \left(R_{\rm sh}\left(t\right)/r_{\rm i}\right)^{-n}V_{\rm sh}\left(t\right)^2 h\left(\xi\right)$ \\
\bottomrule
\label{tab}
\end{tabular}
\end{table*}
Here we analyze perturbations to the self-similar solutions in \citetalias{coughlin23}; we will use their same variables, as summarized in Table \ref{tab:1}. We provide some physical motivation for and a general recapitulation of 
the \citetalias{coughlin23} self-similar solutions, but see \citetalias{coughlin23} for more in-depth discussion.

\subsection{Self-similar solutions}
Immediately prior to the neutrino-induced mass loss event, we assume that at sufficiently large radii the stellar envelope has a density profile that is well-approximated by a power-law in radius and is in hydrostatic balance in a gravitational field dominated by a mass $M$. The initial density and pressure profiles of the stellar envelope are then
\begin{align}
    \rho\left(t=0\right) &= \rho_{\rm i}\left(\frac{r}{r_{\rm i}}\right)^{-n}, \\
    p\left(t=0\right) &= \frac{1}{n+1}\frac{GM\rho_{\rm i}}{r}\left(\frac{r}{r_{\rm i}}\right)^{-n},
\end{align}
where $\rho_{\rm i}$ is the density measured at some scale radius $r_{\rm i}$, $r$ is the radial distance from the central mass, and $n$ is the density power-law index. 

Following the mass loss at $t = 0$, the stellar envelope expands hydrodynamically and the evolution of the fluid velocity $v$, density $\rho$, and pressure $p$ are governed by the continuity, momentum, and entropy equations, which are (in spherical symmetry)
\begin{align}
    \label{cont}
    &\frac{\partial \rho}{\partial t}+\frac{1}{r^2}\frac{\partial}{ \partial r}\left[r^2 \rho v\right]=0, \\
    \label{mom eq}
    &\frac{\partial v}{\partial t}+v\frac{\partial v}{\partial r}+\frac{1}{\rho}\frac{\partial p}{\partial r}=-\frac{GM}{r^2}+\frac{G\delta M}{r^2}, \\
    \label{entropy}
    &\frac{\partial}{\partial t}\ln\left(\frac{p}{\rho^{\gamma}}\right)+v\frac{\partial}{\partial r}\ln\left(\frac{p}{\rho^{\gamma}}\right)=0.
\end{align}
Here $\gamma$ is the adiabatic index of the fluid and 
$M$ has been reduced by $\delta M$ in Equation \eqref{mom eq}. If the mass loss occurs instantaneously, then the only timescale relevant to a given fluid element is its local dynamical time, and hence it is reasonable to assume -- and as was shown to be the case by \citetalias{coughlin23} -- that Equations \eqref{cont}--\eqref{entropy} admit self-similar solutions of the form 
\begin{align}
    \label{ve}
    v_{\rm e} &= \frac{GM}{r}f_{\rm e}\left(\eta\right), \\
    \rho_{\rm e} &= \rho_{\rm i}\left(\frac{r}{r_{\rm i}}\right)^{-n}g_{\rm e}\left(\eta\right), \\
    \label{pe}
    p_{\rm e} &= \frac{1}{n+1}\frac{GM\rho_{\rm i}}{r}\left(\frac{r}{r_{\rm i}}\right)^{-n}h_{\rm e}\left(\eta\right),
\end{align}
where $f_{\rm e}$, $g_{\rm e}$, and $h_{\rm e}$ are functions of the variable 
\begin{align}
    \eta = \frac{\sqrt{GM}t}{r^{3/2}}
\end{align}
and satisfy the hydrostatic initial conditions
\begin{equation}
    f_{\rm e}\left(0\right)=0,\,\, \, g_{\rm e}\left(0\right)=h_{\rm e}\left(0\right)=1.
\end{equation}

The solutions parameterized by Equations \eqref{ve}--\eqref{pe} describe the evolution of the stellar envelope at large radii, i.e., where the fluid is expanding in response to the mass loss and not accreting onto the central object. As one moves to smaller radii at a given time, these solutions terminate in a sonic point, suggesting that this outer region can be joined onto an inner, accreting region via a shockwave with radius $R_{\rm sh}(t)$ that encloses the sonic horizon. \citetalias{coughlin23} showed that if the shockwave satisfies
\begin{align}
    \label{ss shock position}
    \frac{\sqrt{GM}t}{{R_{\rm sh}}^{3/2}}&=\etash, \\
    \Rightarrow\frac{dR_{\rm sh}}{dt}&\equiv V_{\rm sh} = \frac{2}{3\etash}\sqrt{\frac{GM}{R_{\rm sh}}}, \nonumber
\end{align}
where $\etash$ is a constant, then there exist self-similar solutions that achieve this matching, i.e., for $\xi \equiv r/R_{\rm sh}(t) \le 1$ the fluid is described by 
\begin{equation}
\begin{split}
    v &= V_{\rm sh}(t)f_0(\xi) \\
    \rho &= \rho_{\rm i}\left(\frac{R_{\rm sh}(t)}{r_{\rm i}}\right)^{-n}g_0(\xi), \\
    p &= \rho_{\rm i}\left(\frac{R_{\rm sh}(t)}{r_{\rm i}}\right)^{-n}V_{\rm sh}(t)^2h_0(\xi). \label{ps-ss}
\end{split}
\end{equation}
As for the solutions at large radii, the functions $f_0$, $g_0$, and $h_0$ are determined from the fluid equations and the boundary conditions at the shock. However, provided that the relative neutrino mass loss $\delta M/M$ is below an $n$ and $\gamma$-dependent critical value, \citetalias{coughlin23} showed there are \emph{two} such solutions that accrete near the origin and join onto the expanding envelope. The existence of two solutions strongly suggests that one is unstable to small perturbations, which we demonstrate here (we consider the super-critical mass loss case in Section \ref{sec: global}). 

\subsection{Perturbation analysis and eigenvalues}
We impose a small perturbation to the shock position and velocity in Equation \eqref{ss shock position} via 
\begin{align}
    \label{pert shock position}
    \frac{\sqrt{GM}t}{{R_{\rm sh}}^{3/2}}&=\etash+\eta_1\left(\tau\right), \\
    \Rightarrow\frac{dR_{\rm sh}}{dt}&\equiv V_{\rm sh} = \sqrt{\frac{GM}{R_{\rm sh}}}\left[\dot{\eta}_1+\frac{3}{2}\left(\etash + \eta_1 \right)\right]^{-1}, \nonumber
\end{align}
where the perturbation $\eta_1$ is a function of the dimensionless time-like variable
\begin{equation}
    \tau = \ln\left(\frac{R_{\rm sh}}{r_{\rm i}}\right), \label{tau}
\end{equation}
and dots denote derivatives with respect to $\tau$. We similarly define perturbations to the post-shock fluid velocity, density, and pressure by generalizing Equation \eqref{ps-ss} to 
\begin{align}
    \label{ps vel}
    v &= V_{\rm sh}\left[f_0\left(\xi\right)+f_1\left(\xi,\tau\right)\right], \\
    \label{ps rho}
    \rho &= \rho_{\rm i}{\left(\frac{R_{\rm sh}}{r_{\rm i}}\right)}^{-n}\left[g_0\left(\xi\right)+g_1\left(\xi,\tau\right)\right], \\
    \label{ps p}
    p &= \rho_{\rm i}{\left(\frac{R_{\rm sh}}{r_{\rm i}}\right)}^{-n}{V_{\rm sh}}^2\left[h_0\left(\xi\right)+ h_1\left(\xi,\tau\right)\right].
\end{align}
Inserting these definitions into Equations \eqref{cont}--\eqref{entropy} then yields a set of six equations --- three zeroth-order and three first-order --- where the zeroth-order equations are the self-similar equations from \citetalias{coughlin23}. We include all equations in Appendix \ref{sec: local eqs}.

At the shock front ($\xi=1$) the shock jump conditions must be satisfied, where the ambient (pre-shock) quantities are given by the expanding envelope solution. Therefore, to evaluate the ambient quantities at the location of the shock, we let $\eta\rightarrow \etash +\eta_1$ in Equations \eqref{ve}--\eqref{pe}, which upon Taylor-expanding to leading order gives 
\begin{align}
    \label{amb vel}
    v_{\rm e} = &\sqrt{\frac{GM}{R_{\rm sh}}}f_{\rm e}\left(\etash\right)\left[1+\frac{1}{f_{\rm e}\left(\etash\right)} \frac{\partial f_{\rm e}}{\partial \eta}\bigg\lvert_{\etash}\eta_1\left(\tau\right)\right], \\
    \label{amb rho}
    \rho_{\rm e} = &\rho_{\rm i} \left(\frac{R_{\rm sh}}{r_{\rm i}}\right)^{-n}g_{\rm e} \nonumber\\
    &\times\left(\etash\right)\left[1+\frac{1}{g_{\rm e}\left(\etash\right)} \frac{\partial g_{\rm e}}{\partial \eta}\bigg\lvert_{\etash}\eta_1\left(\tau\right)\right], \\
    \label{amb pres}
    p_{\rm e} = &\frac{1}{n+1}\frac{GM}{R_{\rm sh}}\rho_{\rm i} \left(\frac{R_{\rm sh}}{r_{\rm i}}\right)^{-n} \nonumber \\
    &\times h_{\rm e}\left(\etash\right)\left[1+\frac{1}{h_{\rm e}\left(\etash\right)} \frac{\partial h_{\rm e}}{\partial \eta}\bigg\lvert_{\etash}\eta_1\left(\tau\right)\right].
\end{align}
Inserting Equations \eqref{ps vel}--\eqref{ps p} and Equations \eqref{amb vel}--\eqref{amb pres} into the jump conditions, keeping up to first-order terms, and equating terms by order therefore gives the boundary conditions for the unperturbed and perturbed dimensionless fluid quantities, which we provide in Appendix \ref{sec: local eqs}.

Equations \eqref{ss cont}--\eqref{ss ent} can thus be numerically integrated inward from $\xi=1$ to solve for the self-similar post-shock velocity ($f_0$), density ($g_0$), and pressure ($h_0$) profiles for a given ambient density power-law index $n$, adiabatic index $\gamma$, and relative mass loss $\delta M/M$, provided that it is below a maximum value that we denote $\left(\delta M/M\right)_{\rm max}$. It was shown in \citetalias{coughlin23} that --- for a given $n$, $\gamma$, and $\delta M/M$ --- there are two solutions (two values of $\etash$) that smoothly pass through a sonic point in the post-shock flow and accrete. These are the ``strong'' and ``weak'' solutions, with the strong solution having the smaller value of $\etash$ and (therefore) a greater shock velocity and Mach number. 

\begin{figure*}
    \includegraphics[width=0.48\textwidth]{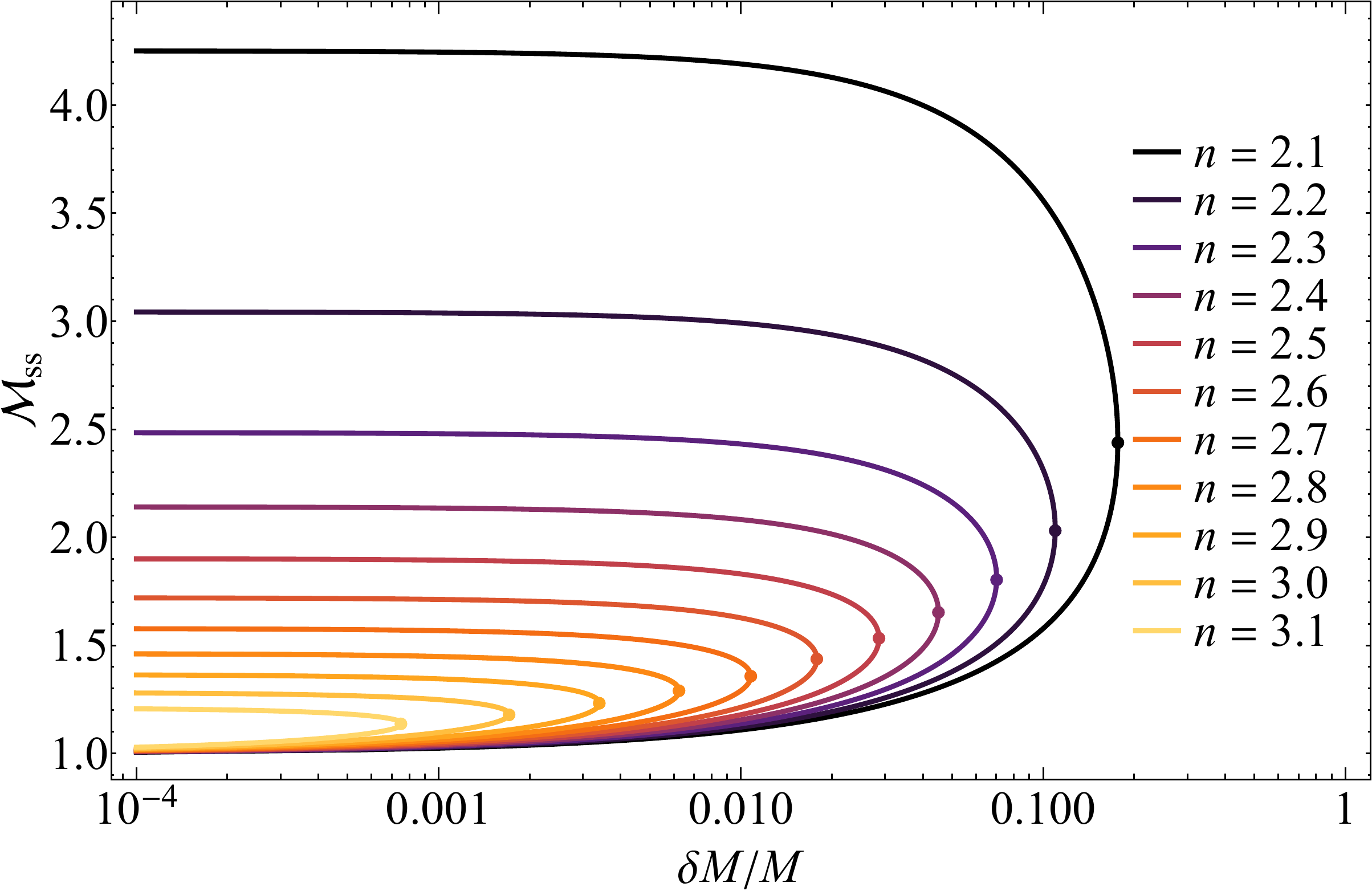}
    \includegraphics[width=0.503\textwidth]{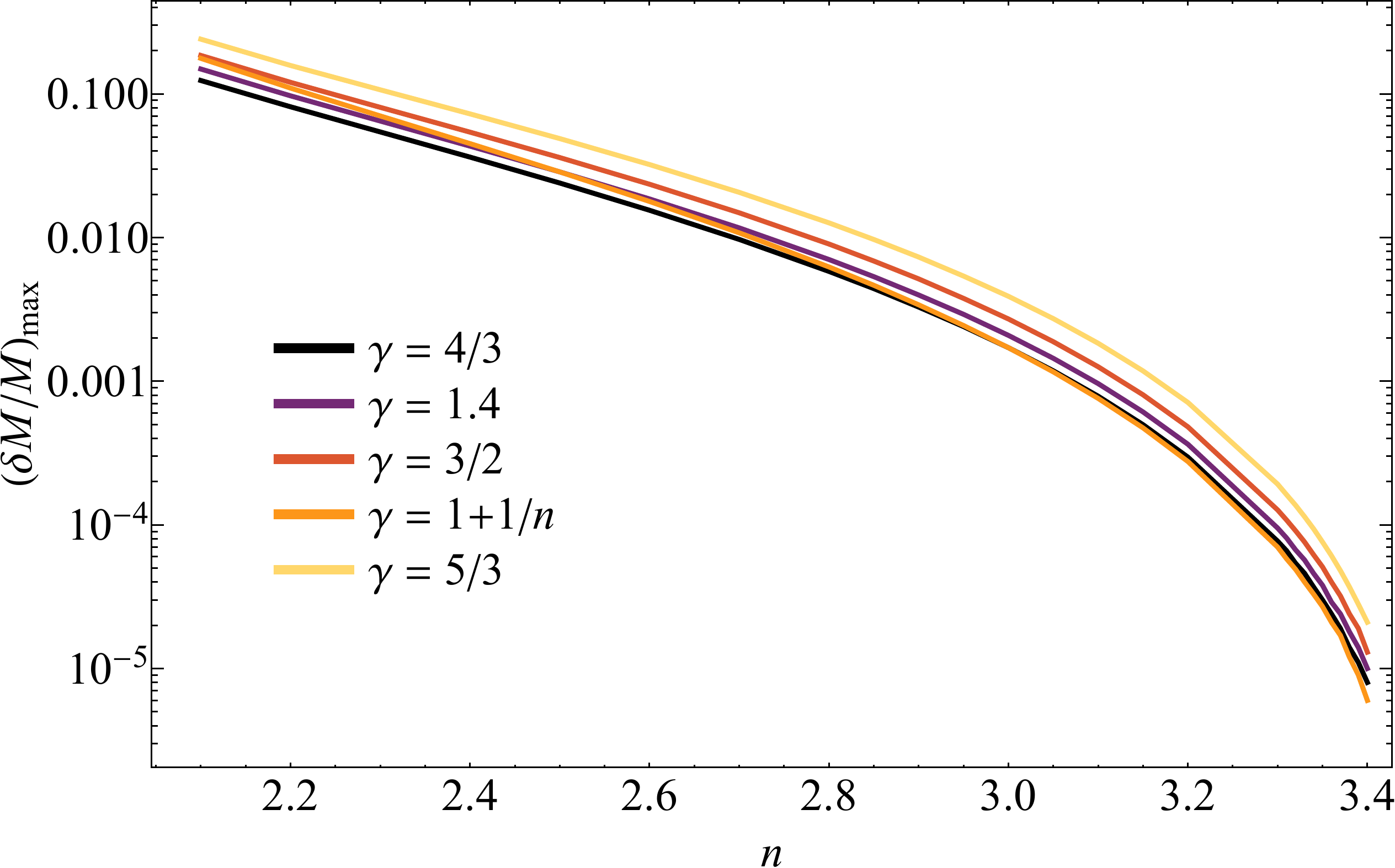}
    \caption{Left: The self-similar shock Mach number as a function of the relative mass loss $\delta M/M$ for the different ambient power-law indices shown in the legend and an adiabatic index of $\gamma=1+1/n$. For a given value of $n$ and $\delta M/M$, there are two solutions for the Mach number, which are the strong (larger Mach number) and weak (smaller Mach number) solutions. As the mass loss increases, the two solutions converge and are equal at a critical mass loss value $\left(\delta M/M\right)_{\rm max}$ (shown by the points) above which no self-similar solution exists. Right: $\left(\delta M/M\right)_{\rm max}$ as a function of the ambient power-law index $n$ for the different adiabatic indices $\gamma$ shown in the legend.}
    \label{fig: dM max}
\end{figure*}

The left panel of Figure \ref{fig: dM max} shows the self-similar shock Mach number as a function of the relative mass loss $\delta M/M$ for a range of power-law indices $n$ and an adiabatic index $\gamma=1+1/n$. For a given $n$, there are two solutions below the corresponding maximum relative mass loss $\left(\delta M/M\right)_{\rm max}$, which is indicated by the point along each curve. As the mass loss nears the maximal value, the strong (larger Mach number) and weak (smaller Mach number) solutions converge and equal one another at $\left(\delta M/M\right)_{\rm max}$. The right panel of the same figure shows $\left(\delta M/M\right)_{\rm max}$ as a function of the ambient power-law index $n$ for the adiabatic indices in the legend. Independent of $\gamma$, it can be seen that as $n$ increases, $\left(\delta M/M\right)_{\rm max}$ decreases and approaches $0$ in the limit that $n\rightarrow3.5$, above which there are no self-similar solutions. 

The perturbed equations, Equations \eqref{pert cont}--\eqref{pert ent}, govern the spatial and temporal evolution of the perturbed post-shock quantities. We recast these equations by taking the Laplace transform in the time-like variable $\tau$, where the Laplace transform of, 
e.g., $f_1$, is
\begin{equation}
    \label{lt f}
    \tilde{f_1}=\int_0^\infty f_1\left(\xi, \tau\right)e^{-\sigma\tau}d\tau.
\end{equation}
We additionally redefine the perturbed fluid quantities by dividing the Laplace Transformed equations by the perturbation $\tilde{\eta}_1$ and letting, for example, $\tilde{f}_1 \rightarrow \tilde{f}_1/\tilde{\eta}_1$. In doing so, and performing the same operations on the boundary conditions, the first-order fluid equations become a system of ordinary differential equations that are independent of the magnitude of the perturbation to the shockwave $\eta_1$, and are given by Equations \eqref{lt pert cont}--\eqref{lt pert ent}. Similar to the zeroth-order equations, $\sigma$ is an ``eigenvalue'' that is determined by requiring that the solutions pass smoothly through the sonic point, and is calculated numerically using a shooting method.

There are infinitely many eigenvalues that satisfy the smoothness of the first-order quantities through the sonic point, and these eigenvalues are generally complex. However, the eigenvalue with the largest real part is also purely real (as also found in previous studies of the stability of expanding blastwaves; e.g., \citealt{Ryu87, Chevalier90, Sari00}), and since it is this eigenvalue that determines the asymptotic behavior of the perturbations, we focus our analysis on these ``lowest-order'' modes. 

\begin{figure*}
    \includegraphics[width=1\textwidth]{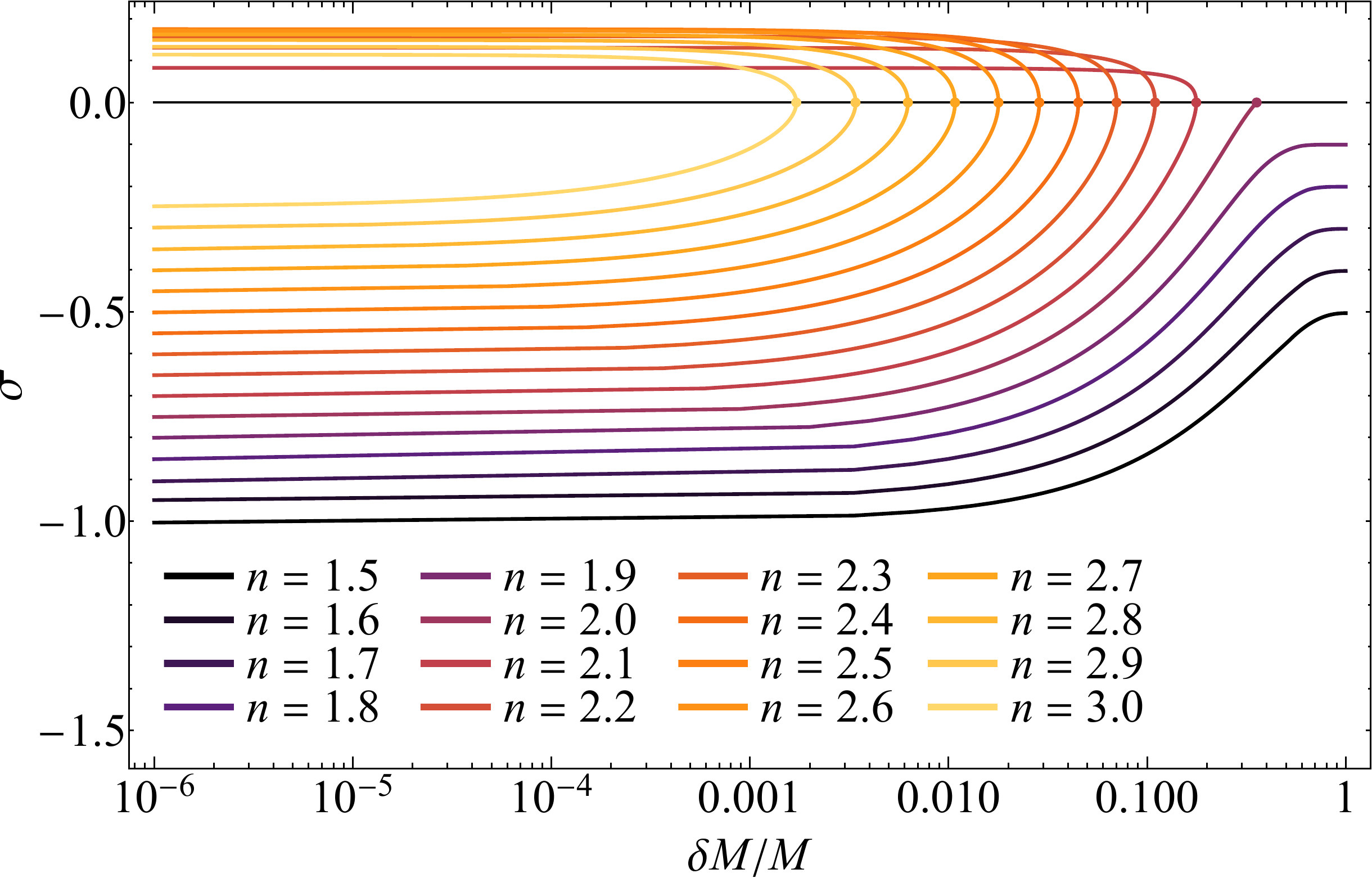}
    \caption{Eigenvalues for the strong (positive, unstable branch) and weak (negative, stable branch) solutions as a function of mass loss for different power-law indices $n$ and adiabatic index $\gamma=1+1/n$. As each solution nears the critical mass loss, the strong and weak shock solutions converge with an eigenvalue of $\sigma=0$. Also shown are solutions with $n\leq2$, which only exist for the weak shock, stable solution.}
    \label{fig: sigma_dM_sub}
\end{figure*}

Figure \ref{fig: sigma_dM_sub} shows the lowest-order eigenvalues for the strong shock and weak shock solutions as a function of the relative mass loss $\delta M/M$ for the range of ambient density power-laws $n$ shown in the legend; here we adopted an adiabatic index of $\gamma=1+1/n$. The positive and negative eigenvalues (for a given mass loss $\delta M/M$) correspond to the strong and weak shock solutions, respectively. Recalling that perturbations to the self-similar solutions scale as $\propto e^{\sigma\tau} \propto {R_{\rm sh}^{\sigma}}$ (Equation \eqref{lt f}), this confirms 
that the strong (weak) shock solution is unstable (stable), and therefore perturbations to the self-similar solutions increase (decrease) with time. As discussed above, for a given $n$ and $\gamma$ there is a critical value of $\delta M/M$ above which no self-similar solution exists, and it can be seen that as the mass loss approaches this maximum value, the strong and weak shock eigenvalues converge and are equal to zero (as indicated by the points along the curves with $n\geq2$). The strong shock self-similar solution does not exist for $n\leq2$, whereas the weak solution does (i.e., for any $n$ there is a stable solution, while there are only unstable solutions for $n > 2$). This figure shows that, in this regime, the weak shock solution exists even in the limit that $\delta M/M\rightarrow1$, with perturbations becoming increasingly stable for shallower power-laws. 

\subsection{Numerical Simulations}
\label{sec: local sim}
\begin{figure}
    \includegraphics[width=0.4725\textwidth]{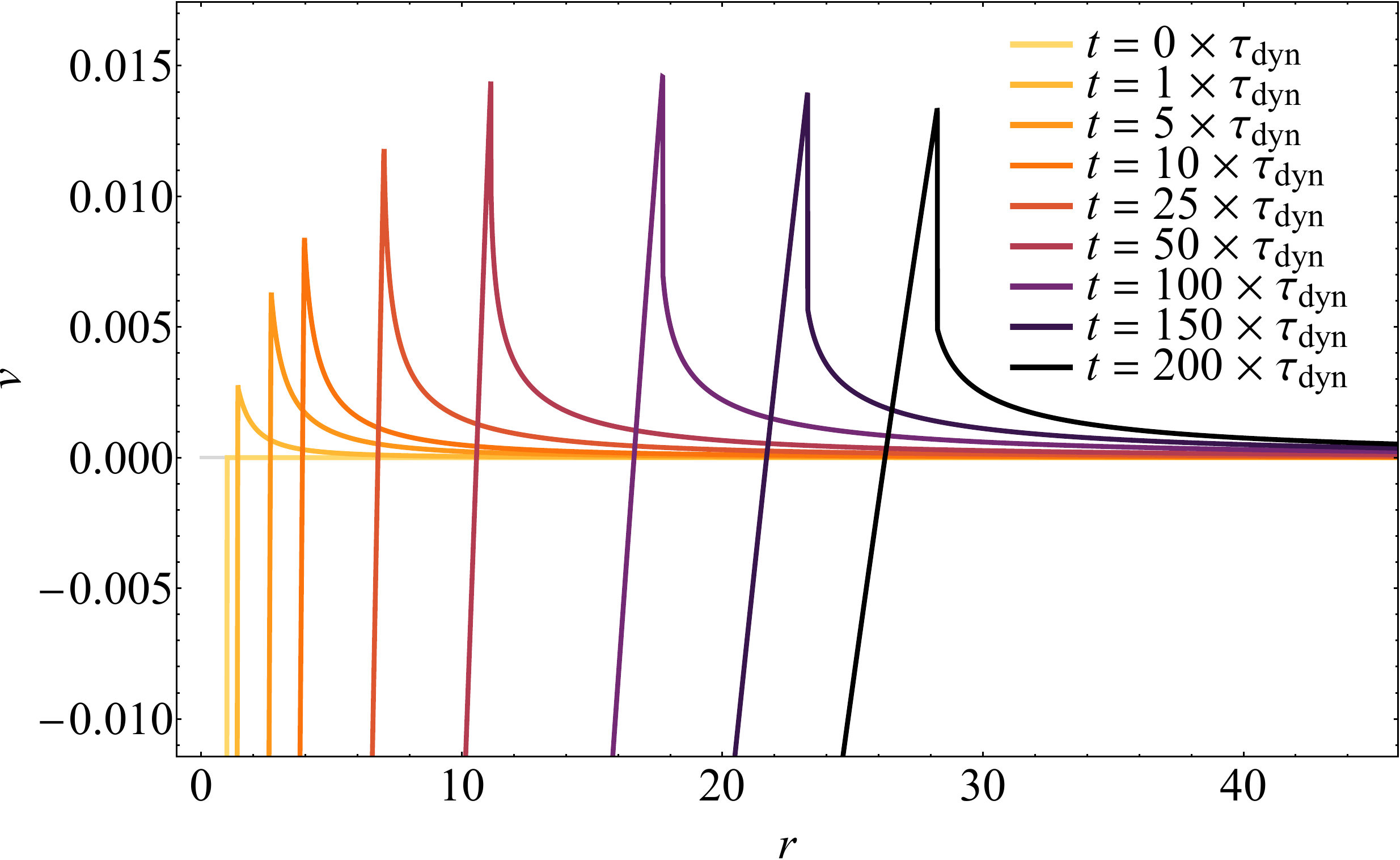}
    \caption{The fluid velocity as a function of radius at different times shown in the legend from a simulation with $n=2.5$, $\gamma=1.4$, and $\delta M/M=0.01$.}
    \label{fig: vr}
\end{figure}

To validate the results of the perturbation analysis presented above, we run $1$-dimensional, uniform grid hydrodynamic simulations with the finite-volume code \textsc{flash} (V$4.7$; \citealt{fryxell00}). We use a resolution of $\Delta r=9.99\times10^{-3}$ and the handling of the boundary conditions and other simulation parameters are the same as those outlined in \citet{Paradiso25}. For each simulation we initialize the domain with the region interior to $r = 1$ given by the rarefaction wave solution of \citealt{coughlin19}, and the region exterior to $r = 1$ is a power-law density profile (for a given power-law index $n$) that is in hydrostatic equilibrium in the gravitational field of a point mass $M$. We start each simulation with the point mass reduced by $\delta M$, reflecting an instantaneous reduction in the mass seeding the gravitational field, which results in the formation of a weak shock. We ran five different simulations with the same relative mass loss value of $\delta M/M=0.01$, power-law indices ranging from $n=2.1$ to $n=2.5$ in steps of $0.1$, and an adiabatic index of $\gamma=1+1/n$. To illustrate the steepening of the rarefaction wave and formation of a shockwave following the mass loss, Figure \ref{fig: vr} shows the fluid velocity as a function of radius at various times for a simulation with $n=2.5$, $\gamma=1.4$, and $\delta M/M=0.01$. Here time is measured with respect to
\begin{equation}
    \label{tau dyn}
    \tau_{\rm dyn}=\frac{{r_{\rm i}}^{3/2}}{\sqrt{GM}},
\end{equation}
which is the dynamical time at the initial location of the rarefaction wave $r_{\rm i}=1$.

\begin{figure*}
    \includegraphics[width=0.496\textwidth]{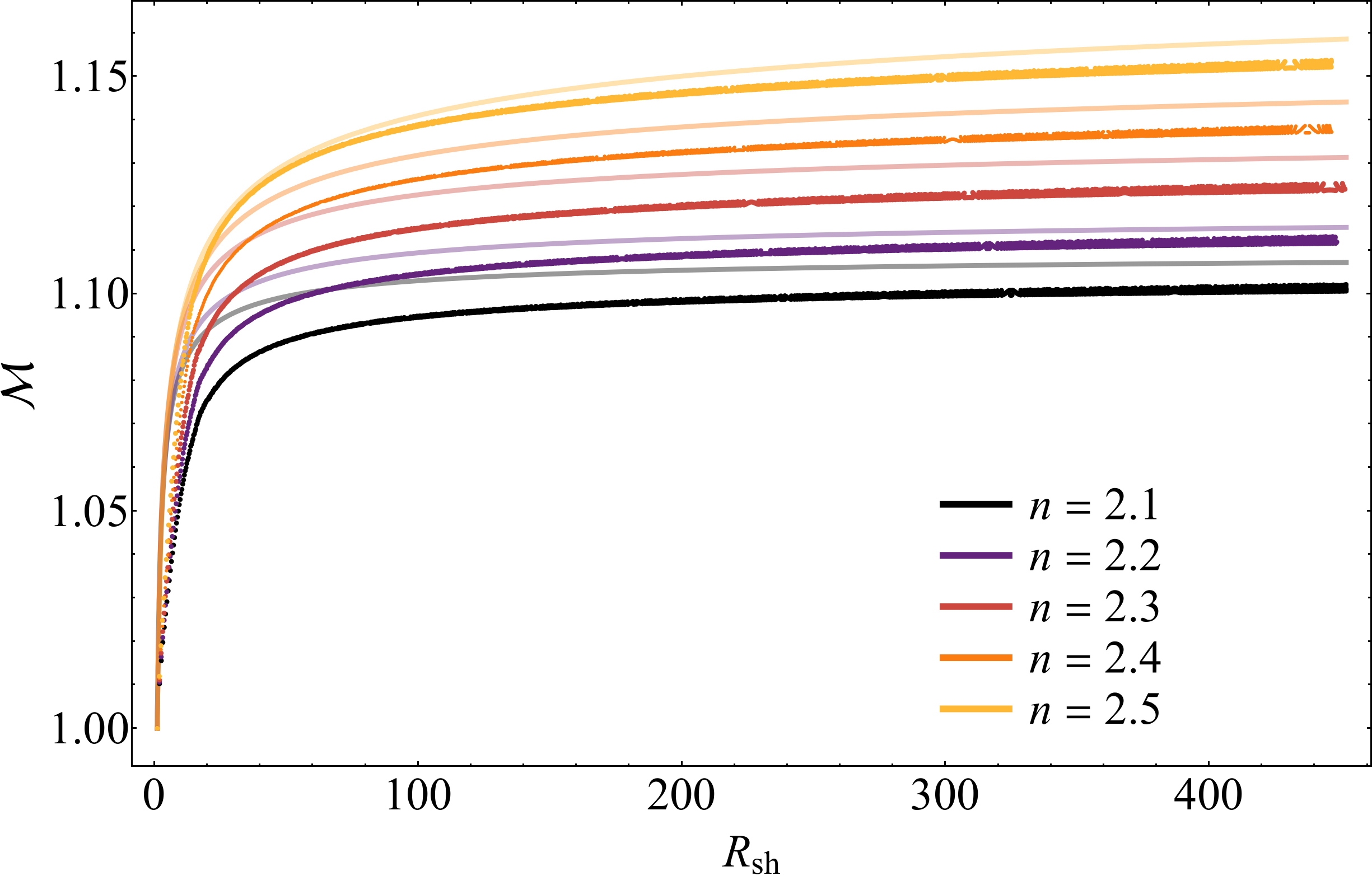}
    \includegraphics[width=0.498\textwidth]{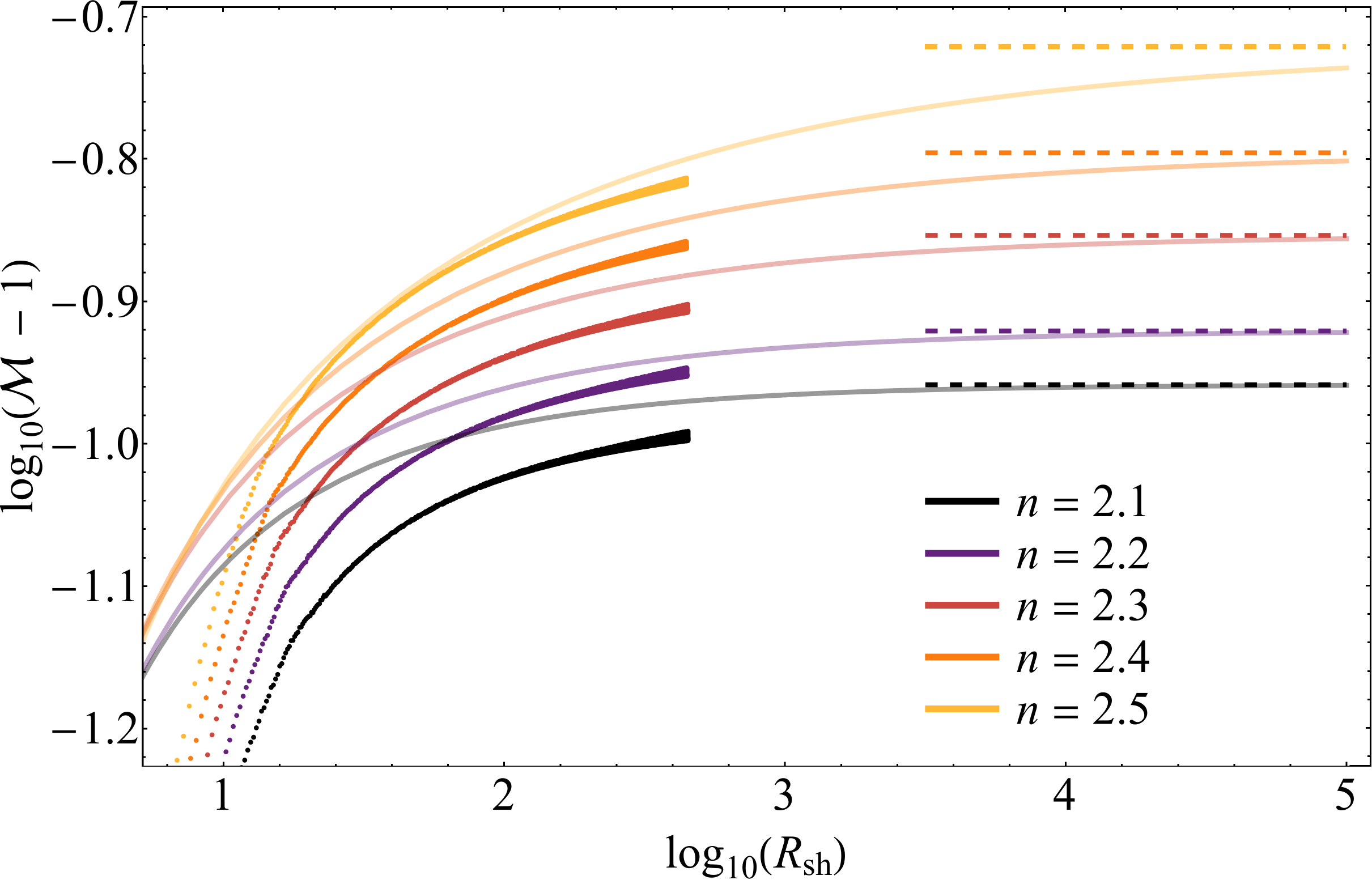}
    \caption{Left: The evolution of the shock Mach number as a function of shock radius measured from the \textsc{flash} simulations (dark dots) for a mass loss of $\delta M/M=0.01$, with ambient density power-law indices $n$ shown in the legend and an adiabatic index of $\gamma = 1+1/n$. The light, solid lines show the approximate evolution of the Mach number as predicted by Equation \ref{Mach evo} and the appropriate value of $\sigma$ given in Table \ref{tab:2}. Right: The log of the difference between the time-evolving Mach number and $1$ as a function of the log of the shock radius, with the analytic curves (light solid) extended to larger radii. Here the horizontal dashed lines indicate the corresponding self-similar value, which is the expected value each simulation will converge to.}
    \label{fig: Mach}
\end{figure*}

Once it forms, the temporal evolution of the shock Mach number can be well approximated by
\begin{equation}
    \label{Mach evo}
    \mathcal{M}\left(R_{\rm sh}, t\right) = \mathcal{M}_{\rm ss} - \left(\mathcal{M}_{\rm ss}-1\right)\left(\frac{R_{\rm sh}}{r_{\rm i}}\right)^{\sigma},
\end{equation}
where $\mathcal{M_{\rm ss}}$ is the asymptotic self-similar Mach number 
and $\sigma$ is the eigenvalue from the perturbation analysis (Table \ref{tab:2} gives a list of $\mathcal{M}_{\rm ss}$ and $\sigma$ for $\delta M/M=0.01$). The left panel of Figure \ref{fig: Mach} shows the Mach number as a function of shock radius for the results of our five numerical simulations (dark dots) and the analytical prediction from Equation \eqref{Mach evo} (light curves). The right panel illustrates the log of the difference between the shock Mach number and $1$ as a function of the log of the shock position. This representation highlights the discrepancies between the numerical and analytical solutions at early times (small shock radii), but these differences are largely irrelevant because the Mach number is effectively one, and shows the trend of the two solutions converging over time. However, while it appears that the $n=2.5$ numerical and analytical solutions converge and then deviate away from one another at late times, this is due to the assumption in Equation \eqref{Mach evo} that the amplitude of the perturbation grows as the difference between the self-similar Mach number and $1$, thereby over-predicting the perturbation contribution at early times. In this plot, we extend the analytical curves to larger radii to demonstrate the slow growth of these perturbations. We additionally indicate the asymptotic values, $\log_{10}\left(\mathcal{M}_{\rm ss}-1\right)$, by the horizontal dashed curves. 

\begin{deluxetable*}{lccccc}
\tablecaption{The Self-Similar Mach Number $\mathcal{M}_{\rm ss}$, Maximum Relative Mass Loss $\left(\delta M/M\right)_{\rm max}$, and Growth Rate $\sigma$ at Which the Numerical Solutions Converge to the Self-Similar Value for $\delta M/M=0.01$ and Ambient Density Power-Law Indices $n$ Shown in the Left Column.}
\label{tab:2}
\tablewidth{0pt}
\tablehead{
  \colhead{$\{\mathcal{M}_{\rm ss}, \left(\delta M/M\right)_{\rm max}, \sigma\}$} &
  \colhead{$\gamma = 1 + 1/n$} &
  \colhead{$\gamma = 4/3$} &
  \colhead{$\gamma = 1.4$} &
  \colhead{$\gamma = 1.5$} &
  \colhead{$\gamma = 5/3$}
}
\startdata
$n = 2.1$ & $\{1.11, 0.177, -0.596\}$ & $\{1.12, 0.124, -0.567\}$ & $\{1.12, 0.149, -0.582\}$ & $\{1.11, 0.185, -0.599\}$ & $\{1.09, 0.240, -0.620\}$\\
$2.2$ & $\{1.12, 0.110, -0.526\}$ & $\{1.14, 0.0812, -0.498\}$ & $\{1.13, 0.0968, -0.515\}$ & $\{1.12, 0.120, -0.535\}$ & $\{1.10, 0.158, -0.558\}$\\
$2.3$ & $\{1.14, 0.0702, -0.454\}$ & $\{1.15, 0.0543, -0.426\}$ & $\{1.14, 0.0647, -0.445\}$ & $\{1.13, 0.0805, -0.468\}$ & $\{1.11, 0.107, -0.494\}$\\
$2.4$ & $\{1.16, 0.0450, -0.377\}$ & $\{1.17, 0.0363, -0.351\}$ & $\{1.16, 0.0433, -0.373\}$ & $\{1.14, 0.0541, -0.398\}$ & $\{1.13, 0.0725, -0.428\}$\\
$2.5$ & $\{1.18, 0.0286, -0.294\}$ & $\{1.19, 0.0240, -0.268\}$ & $\{1.18, 0.0286, -0.294\}$ & $\{1.16, 0.0360, -0.325\}$ & $\{1.14, 0.0488, -0.360\}$\\
$2.6$ & $\{1.21, 0.0178, -0.199\}$ & $\{1.23, 0.0155, -0.172\}$ & $\{1.21, 0.0186, -0.206\}$ & $\{1.19, 0.0235, -0.244\}$ & $\{1.16, 0.0322, -0.286\}$\\
$2.7$ & $\{1.29, 0.0108, -0.0611\}$ & $\{--, 0.00970, --\}$ & $\{1.27, 0.0117, -0.0900\}$ & $\{1.22, 0.0149, -0.149\}$ & $\{1.19, 0.0206, -0.205\}$\\
\enddata
\end{deluxetable*}

\section{Global Stability}
\label{sec: global}
As discussed in Section \ref{sec: intro}, the self-similar solutions 
exist if $\delta M/M$ is below a critical value, $\left(\delta M/M\right)_{\rm max}$, which is shown in the right panel of Figure \ref{fig: dM max} as a function of $n$ for various $\gamma$. We can understand the super-critical evolution by including the excess mass loss $\Delta M/M$ (i.e., the total mass lost minus the critical mass loss) as a driving term that excites perturbations to the critical-mass-loss self-similar solutions. More specifically and analogously to the approach used in Section \ref{sec: local}, we let
\begin{equation}
    \label{eta sup crit}
    \frac{\sqrt{GM}t}{{R_{\rm sh}^{3/2}}}=\eta_{\rm sh}+\frac{\Delta M}{M}\eta_1\left(\tau\right),
\end{equation}
where again $\tau$ is the log of the shock position, $\Delta M/M$ is the excess mass loss for which the self-similar solutions cannot account, i.e.,
\begin{equation}
    \frac{\Delta M}{M} +\left(\frac{\delta M}{M}\right)_{\rm max}
\end{equation}
is the total mass lost. The shock position and velocity are therefore identical to Equation \eqref{pert shock position} with all subscript-$1$ quantities $\propto \Delta M/M$.

We also write the ambient fluid variables in terms of their unperturbed (such that the mass loss is equal  to the critical value) and perturbed (accounting for the excess mass loss above the critical value) components, i.e., we let 
\begin{align}
    \label{ve sup crit}
    v_{\rm e} &= \sqrt{\frac{GM}{r}}\left[f_{\rm e,0}\left(\eta\right)+\frac{\Delta M}{M} f_{\rm e,1}\left(\eta\right)\right], \\
    \rho_{\rm e} &= \rho_{\rm i}\left(\frac{r}{r_{\rm i}}\right)^{-n}\left[g_{\rm e,0}\left(\eta\right)+\frac{\Delta M}{M} g_{\rm e,1}\left(\eta\right)\right], \\
    \label{pe sup crit}
    p_{\rm e} &= \frac{1}{n+1}\frac{GM}{r}\rho_{\rm i}\left(\frac{r}{r_{\rm i}}\right)^{-n}\left[h_{\rm e,0}\left(\eta\right)+\frac{\Delta M}{M} h_{\rm e,1}\left(\eta\right)\right].
\end{align}

\begin{figure}
    \includegraphics[width=0.4725\textwidth]{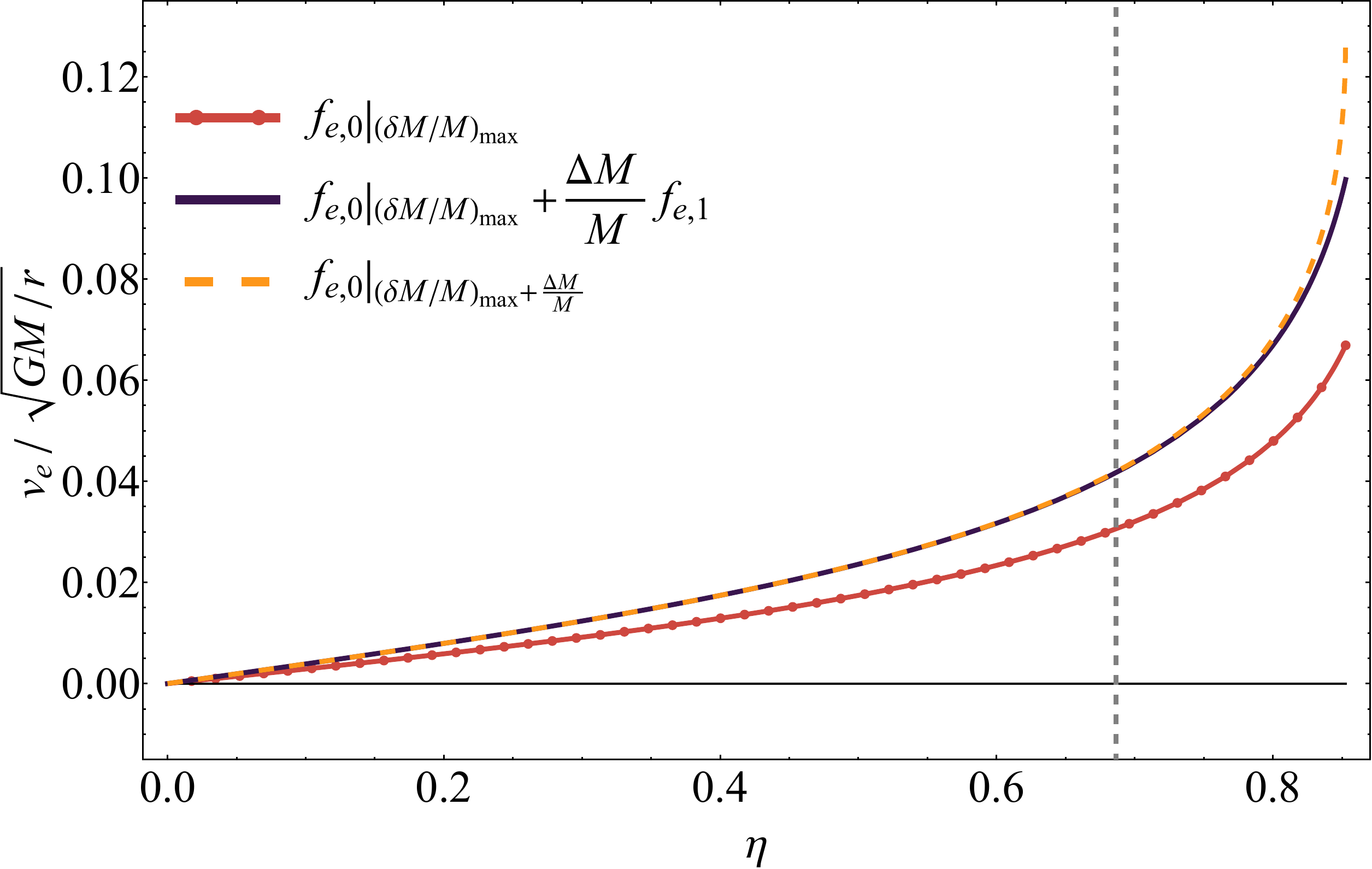}
    \caption{The dimensionless ambient velocity as a function of the dimensionless dynamical time $\eta=\sqrt{GM}t/r^{3/2}$, for $n=2.5$, $\gamma=1.4$, and $\Delta M/M=0.01$. Since the ambient solutions exist in the super-critical max loss regime, we show the solution for the maximum mass loss with the red, dotted curve, the solution at the total mass loss (i.e., solving the zeroth order ambient equations with a mass loss of $\left(\delta M/M\right)_{\rm max}+\Delta M/M$) with the yellow, dashed curve, and the perturbed solution given by Equation \eqref{ve sup crit} with the dark blue curve. The vertical dashed line indicates the value of $\etash$ for $n=2.5$, $\gamma=1.4$, and $\left(\delta M/M\right)_{\rm max}$, which is the ``location'' that the ambient solutions are evaluated at.}
    \label{fig: ve}
\end{figure}

Plugging Equations \eqref{ve sup crit}--\eqref{pe sup crit} into the fluid equations (Equations \eqref{cont}--\eqref{entropy}) then gives a set a zeroth- and first-order equations, the latter of which we include in Appendix \ref{sec: global eqs}. Since the ambient solution exists in the super-critical mass loss regime, we compare the dimensionless ambient velocity solutions for the unperturbed (maximum mass loss) solution (red, dotted curve), the perturbed solution (dark blue, solid curve), and the ``true,'' total mass loss solution (yellow, dashed curve) for $n=2.5$, $\gamma=1.4$, and $\Delta M/M=0.01$ in Figure \ref{fig: ve}. We also indicate the weak-shock $\etash$ by the vertical dashed line. 

We also write the post-shock quantities in the same form as Equations \eqref{ps vel}--\eqref{ps p} but let all subscript-$1$ quantities be $\propto\Delta M/M$. The only difference between the eigenmode analysis of the preceding section and the current analysis is therefore the driving term in the momentum equation: the deviation from the self-similar solution is imparted by the excess mass lost to neutrinos, not differences in the initial conditions -- as would be the case for the eigenvalues. We are therefore not interested in the (already-known from the previous section) eigenvalues themselves, but in the coefficient that multiplies the most-unstable term in the eigenvalue expansion of the background (self-similar) solution. Since this driving term is time-independent and the largest (in an absolute-value sense) eigenvalue appropriate to the maximum-mass-loss self-similar solutions is $\sigma = 0$, the unstable perturbation to the shock position imparted by the excess mass loss grows as
\begin{equation}
    \eta_1 = c_1 \tau = c_1\ln\left(\frac{R_{\rm sh}}{r_{\rm i}}\right), \label{eta1eq}
\end{equation}
where $c_1$ is the coefficient that we want to determine; this can be achieved numerically from the behavior of the Laplace-transformed shock perturbation near the pole at $\sigma = 0$ (see \citealt{Coughlin20} and Appendix \ref{sec: global eqs}). The corresponding shock position and velocity are then
\begin{equation}
    \label{sup crit eta}
    \frac{\sqrt{GM}t}{R_{\rm sh}^{3/2}} = \eta_{\rm sh}+\frac{\Delta M}{M}c_1\tau,
\end{equation}
\begin{equation}
    \label{Vsh sup crit}
    V_{\rm sh} = \sqrt{\frac{GM}{R_{\rm sh}}}\left[\frac{\Delta M}{M}c_1+\frac{3}{2}\left(\etash+\frac{\Delta M}{M}c_1 \tau\right)\right]^{-1}.
\end{equation}
Note that in both of these expressions, the correction induced by the excess mass loss is assumed to be small, and hence the shock position can be written explicitly (and self-consistently) as
\begin{multline}
    \label{R approx}
    R_{\rm sh}(t) = \left(\frac{\sqrt{GM}t}{\eta_{\rm sh}}\right)^{2/3} \\ 
    \times\left(1-\frac{4}{9}\frac{\Delta M}{M}\frac{c_1}{\eta_{\rm sh}}\ln\left(\frac{\sqrt{GM}t}{\eta_{\rm sh}r_{\rm i}^{3/2}}\right)\right).
\end{multline}
An analogous expression follows for the velocity. The excess mass loss should drive the shock out at a faster rate than would otherwise be the case, and hence we expect $c_1 < 0$.

\begin{figure}
    \includegraphics[width=0.4725\textwidth]{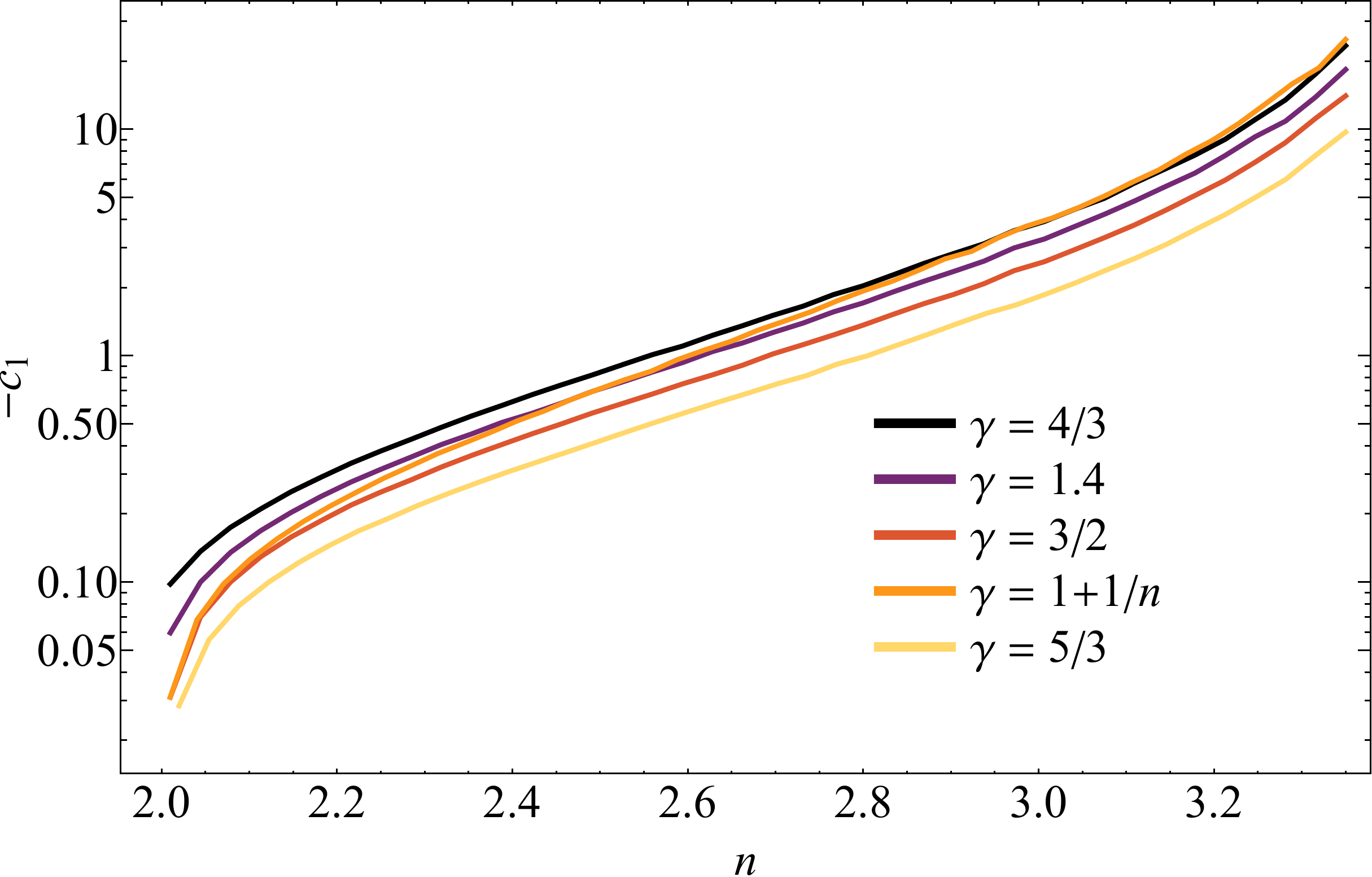}
    \caption{The negative of the constant pre-factor to the logarithmically growing perturbations defined by Equation \eqref{eta1eq} as a function of power-law index $n$ for the different adiabatic indices $\gamma$ shown in the legend. For all values of $\gamma$, $c_1$ becomes increasingly negative for larger $n$, and is $\gtrsim1$ for $n\gtrsim2.9$.}
    \label{fig: c1}
\end{figure}
Figure \ref{fig: c1} shows $-c_1$ as a function of the ambient power-law index $n$ for the adiabatic indices $\gamma$ shown in the legend. We see that $c_1$ becomes increasingly negative for larger $n$, with $|c_1| \gtrsim 1$ for $n\gtrsim2.9$. Physically, the value of $\Delta M/M$ is dependent on the time taken for the star to reach the Tolman-Oppenheimer-Volkov (TOV) limit, which itself depends on the (poorly constrained) neutron star equation of state. Nonetheless, for absolute mass losses of the order $0.5 M_{\odot}$ and less, most systems should have $\Delta M/M$ not well in excess of the critical value (see Tables 1 \& 2 in \citealt{Fernandez18}), and hence -- because $|c_1| \lesssim 1$ for most regions of parameter space -- this linear regime is likely applicable to most progenitors; see Section \ref{sec: implications} for additional discussion.

\subsection{Numerical Simulations}
We tested the accuracy of the above analysis with additional 1-D {\sc flash} simulations. Because the growth of the perturbations is extremely weak, we initialized the simulations with the self-similar shockwave solution near the maximum mass loss, which contrasts the approach of the preceding section that started with a rarefaction wave and hydrostatic envelope. Specifically, we set the initial location of the shock to be $R_{\rm sh} = r_{\rm i} = 1$ and let the fluid variables interior to $r_{\rm i}$ be given by the maximum mass loss solution for $n=2.5$ and $\gamma=1+1/n=1.4$ (being $\left(\delta M/M\right)_{\rm max}=0.0286$). The fluid variables at $r>r_{\rm i}$ are initialized with the ambient solution for the same $n$ and $\gamma$ but at the \textit{total} mass loss. For example, simulations with an excess mass loss of $\Delta M/M=0.01$ adopt the ambient ($r > r_{\rm i}$) self-similar solution with mass loss of $0.0286+\Delta M/M=0.0386$, and we change the point mass value in \textsc{flash} (see Section \ref{sec: local sim}) by this same value. We again use a grid with $10^5$ cells with outflow and reflecting boundary conditions at the inner and outer boundaries at $r_{\rm in} = 0.1$ and $r_{\rm out} = 500$, respectively. 

\begin{figure}
    \includegraphics[width=0.4725\textwidth]{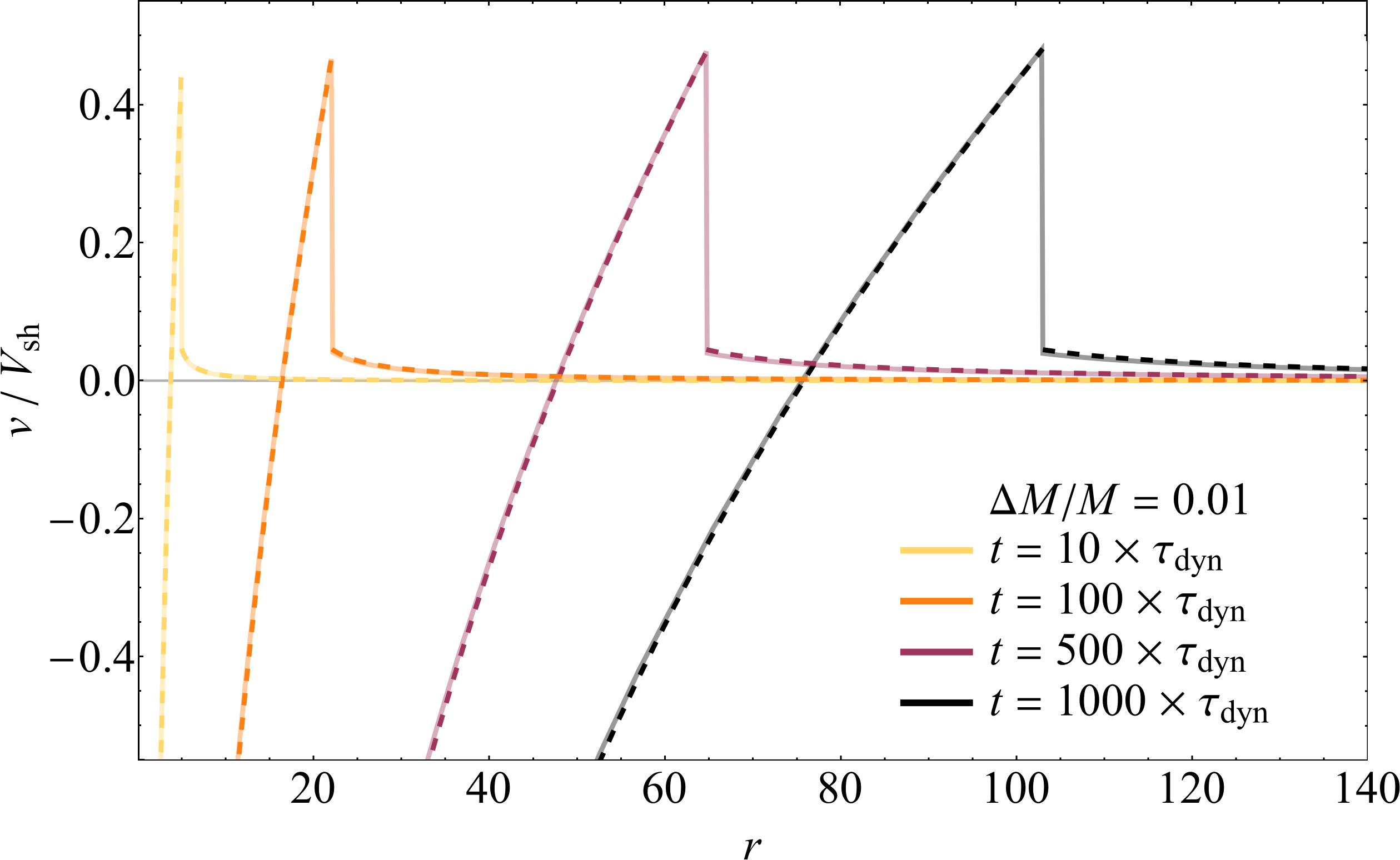}
    \caption{The radial fluid velocity profile normalized by the (time-dependent) shock velocity from the numerical \textsc{flash} simulation (light, solid) and the analytical perturbed solution (dark, dashed) for $n=2.5$, $\gamma=1.4$, and $\Delta M/M=0.01$ shown at different times indicated in the legend.}
    \label{fig: v_comp_sup_crit}
\end{figure}

We performed three simulations with $n = 2.5$, $\gamma = 1.4$, and mass losses $\Delta M/M = 0$, $0.005$, and $0.01$; the $\Delta M/M = 0$ case effectively serves as a control and should maintain a constant Mach number of $\mathcal{M} \simeq 1.52$ if the self-similar solution is accurate. Figure \ref{fig: v_comp_sup_crit} shows the fluid velocity normalized by the time-dependent shock velocity as a function of radius for the $\Delta M/M=0.01$ simulation at the four different times in the legend. Here the light, solid curves are from the numerical simulations, while the dashed curves are the analytical predictions; it is clear that the two are in near-exact agreement. 

\begin{figure*}
    \includegraphics[width=0.496\textwidth]{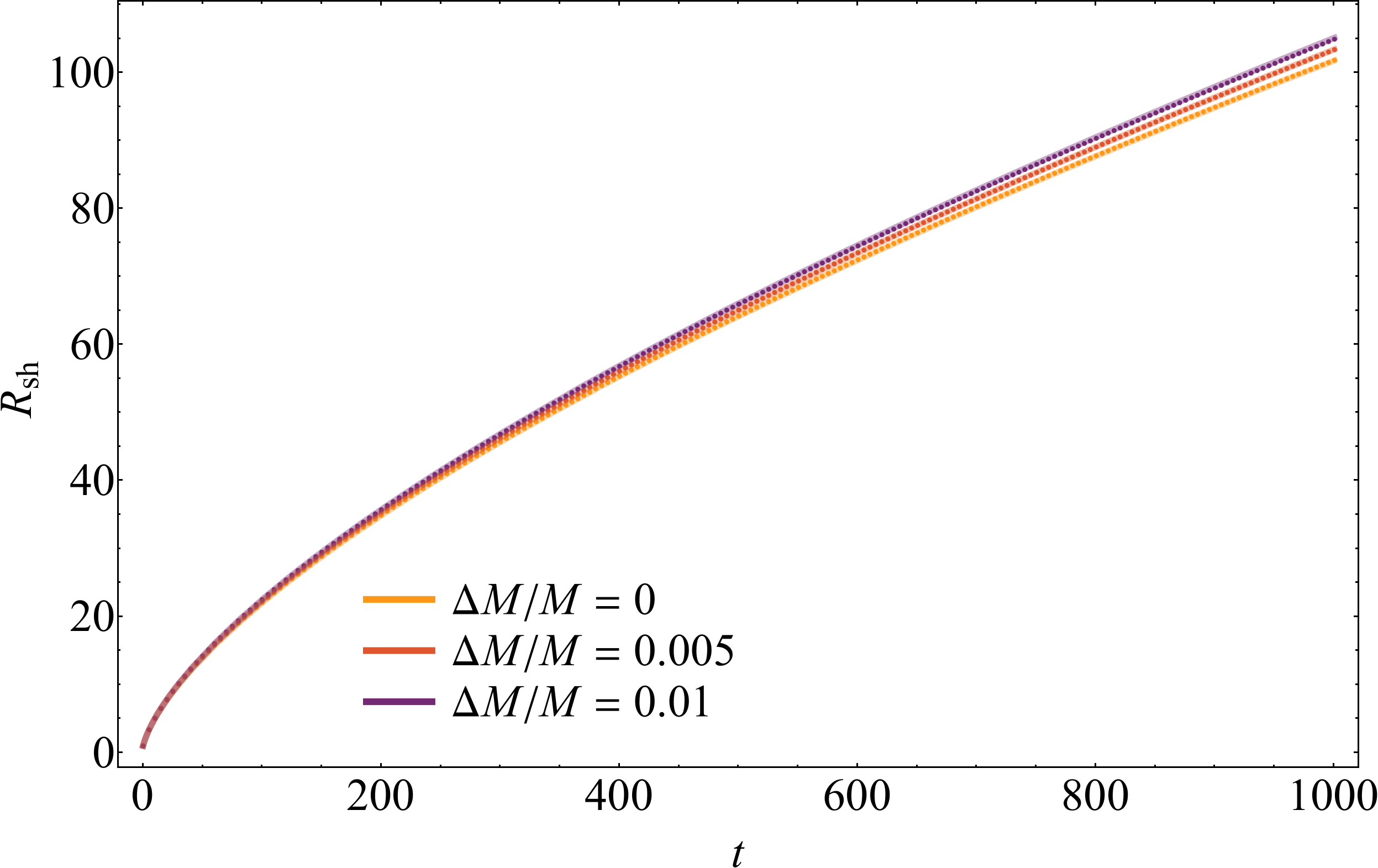}
    \includegraphics[width=0.498\textwidth]{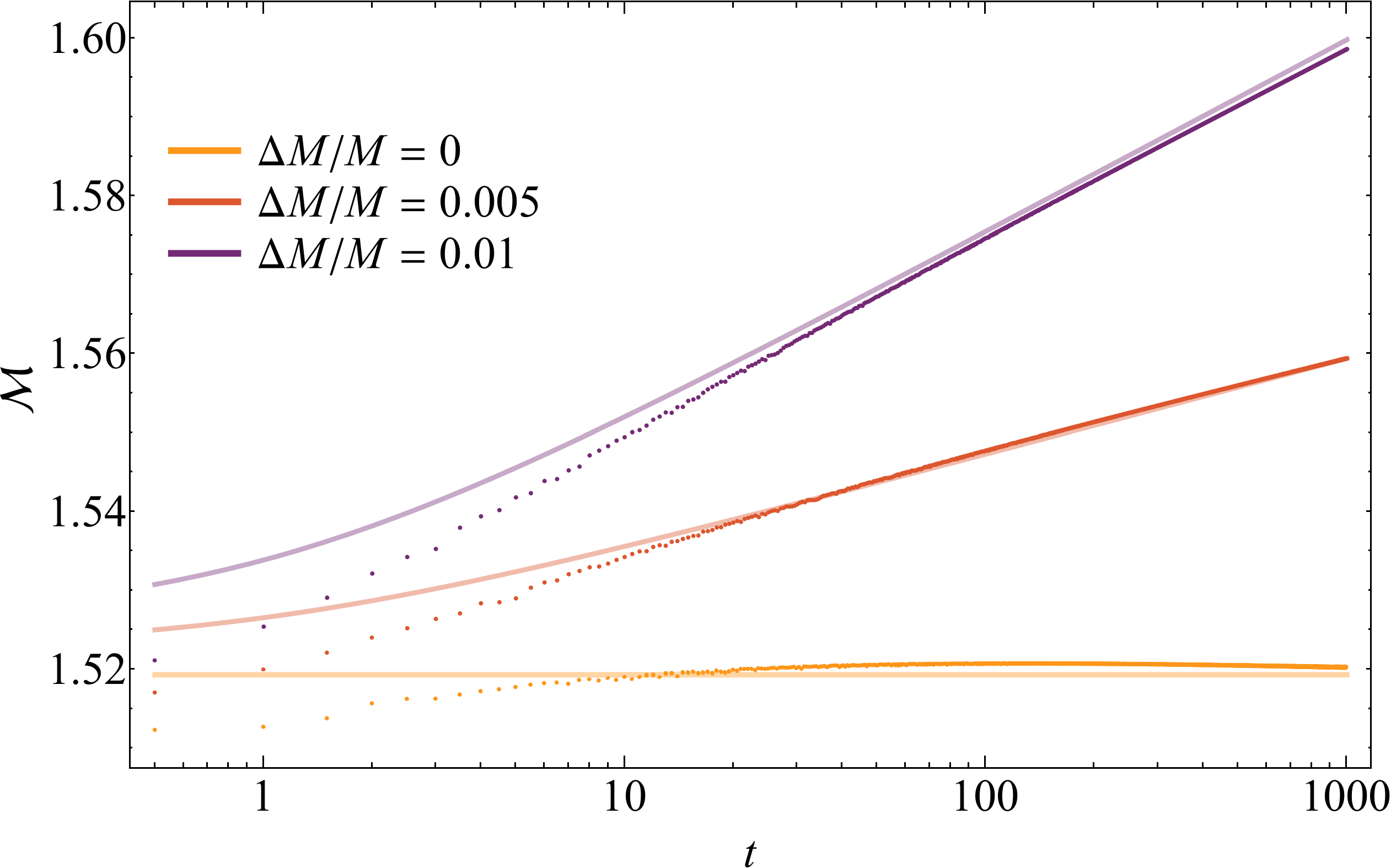}
    \caption{Left: The shock position as a function of time for $n=2.5$ and $\gamma=1.4$ from three different \textsc{flash} numerical simulations (dark points) --- each with different mass losses shown in the legend --- and the analytical prediction (light, solid). By definition --- i.e., Equation \eqref{eta sup crit} --- the solution with $\Delta M/M=0$ is given by the self-similar solution. Right: Comparison of the analytical (light, solid) and numerical (dark points) shock Mach number as a function of time.}
    \label{fig: R shock}
\end{figure*}

The left panel of Figure \ref{fig: R shock} shows the shock trajectories as a function of time for the three different simulations: the numerical results are given by the dark dots and the analytical predictions --- from Equation \eqref{eta sup crit} --- are shown as light solid lines. Here time is again measured with respect to the dynamical time at the initial location of the shock (see Equation \ref{tau dyn}). From the left panel of this figure we see that the analytical and numerical solutions show exceptional agreement, and, as expected, the solutions with non-zero $\Delta M/M$ out-pace the self-similar ($\Delta M/M=0$) solution. To better illustrate this divergence and the accuracy of the linear perturbation analysis, the right panel of this figure shows the shock Mach number as a function of time. As our simulation setup is an amalgamation of the critical mass loss post-shock solution and the super-critical mass loss ambient medium solution, the simulation profiles exhibit a period of relaxation before settling along the corresponding analytical curves. In agreement with the left panel, it can be seen that the analytical curves very accurately predict the growth of the shock Mach number and the deviation away from the $\left(\delta M/M\right)_{\rm max}$ self-similar solution. 

\section{Implications for Real Stars and Producing Strong Explosions}
\label{sec: implications}
There are three important implications of the preceding results in the context of FSNe: 
1) large values of the {relative} mass loss, $\delta M/M(r_{\rm i})$, lead to stronger (larger Mach number) shockwaves, where $M(r_{\rm i})$ is the mass enclosed at the radius $r_{\rm i}$ -- the radius coincident with the head of the sound sound wave during a core-collapse event when the neutrino mass loss occurs; 2) ambient densities that decline faster with radius are -- for the same relative mass loss -- more readily capable of generating strong explosions, as the critical mass loss is a declining function of the ambient power-law index (see Figure \ref{fig: dM max}); and 3) for mass losses that are only marginally above the $n$-$\gamma$-dependent critical value, the shock strengthens only logarithmically with radius, i.e., the Mach number is effectively constant and equal to a value of the order unity from the time of shock formation to the time of shock breakout.  

Given these conclusions, one would naively propose that no massive star should ever generate an energetic explosion, as most stars at the time of core collapse have total masses $\sim 10-15 M_{\odot}$, the absolute mass lost to neutrinos is at most $0.5 M_{\odot}$, and the power-law index of the hydrogen envelope is not very steep (between $r^{-2\pm 0.5}$); i.e., the mass loss is at most only modestly in excess of the critical mass loss. However, Fernández et al. (\citeyear{Fernandez18}; hereafter \citetalias{Fernandez18}) investigated FSNe for a range of stellar progenitors, and found that red supergiants (RSGs) could eject most of their hydrogen envelope in a strong explosion (see also Figure 13 in \citealt{coughlin18}, which demonstrates that the shockwave in a RSG progenitor following the mass loss is in the Sedov phase). The hydrogen envelopes of RSGs also have shallow density profiles in comparison to more compact progenitors, yet the FSNe from those compact progenitors -- blue (BSGs), yellow supergiants (YSGs), and Wolf-Rayets -- were found by \citetalias{Fernandez18} to yield substantially less energetic explosions, ejecting little to no mass. These findings therefore seem to contradict the expectations based on the preceding analyses. 

\begin{figure*}
    \includegraphics[width=0.5\textwidth]{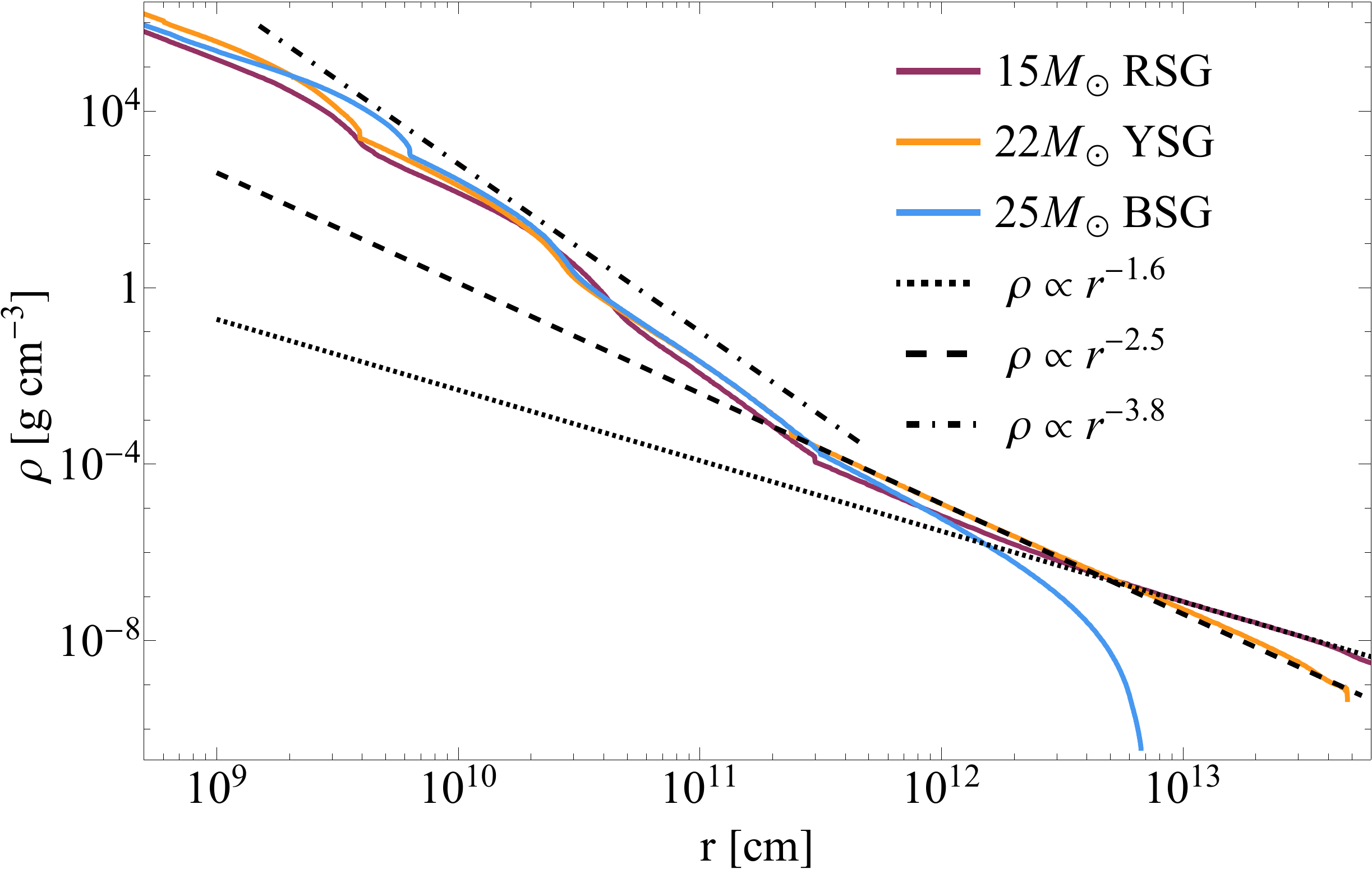}
    \includegraphics[width=0.495\textwidth]{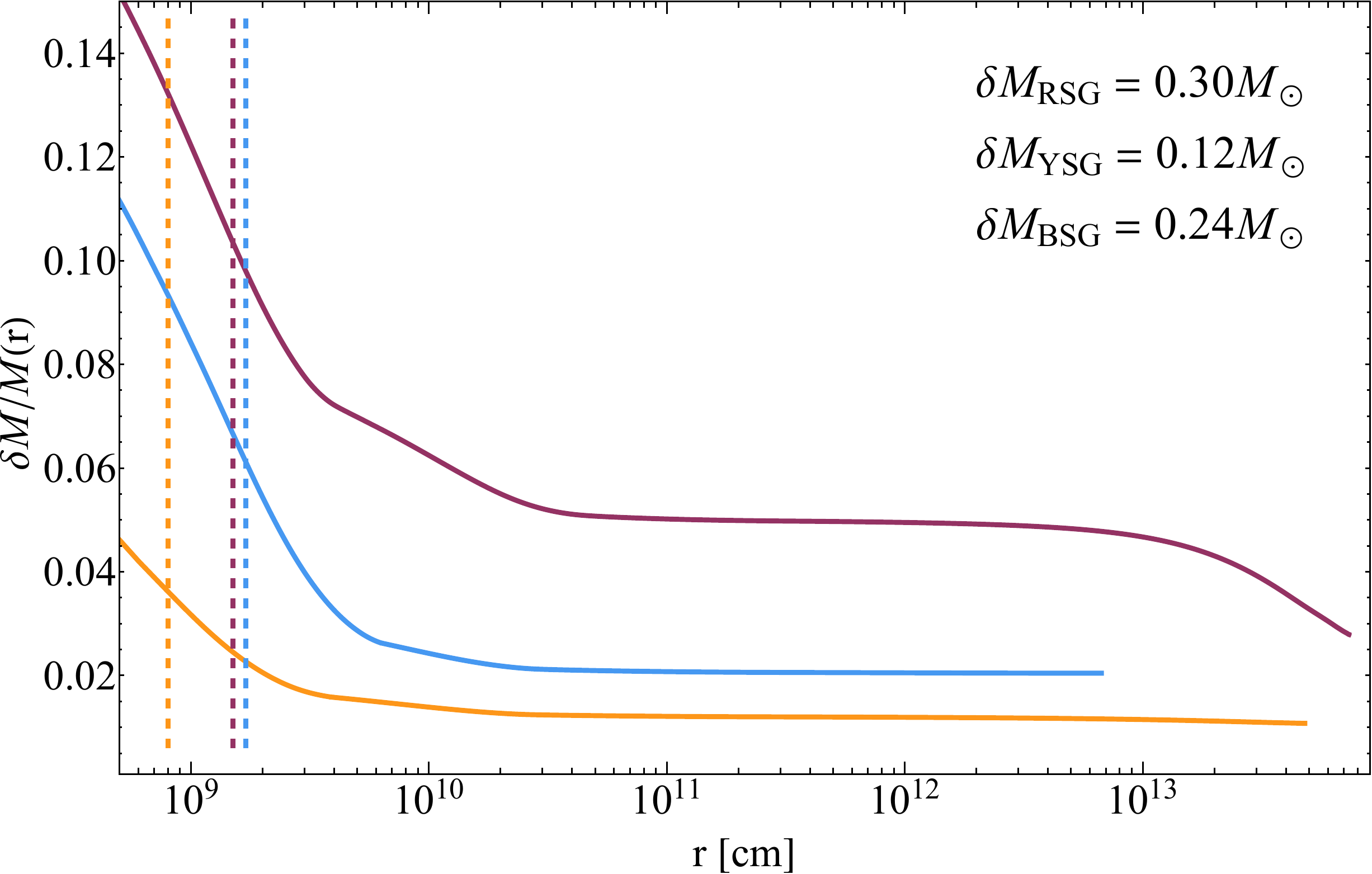}
    \caption{Left: Density as a function of radius for the $15\Msun$ zero age main sequence RSG, $22\Msun$ YSG, and $25\Msun$ BSG progenitors of \citetalias{Fernandez18} at core collapse. Here the dotted and dashed curves show the approximate scalings of the RSG and YSG hydrogen envelopes, which are $\rho \propto r^{-1.6}$ and $\propto r^{-2.5}$, respectively. The dot-dashed curves shows the approximate density decline from the inner region of the star to the base of the hydrogen envelope. Right: The fractional mass loss as a function of radius for the three progenitors, where the vertical lines indicate the location that the rarefaction wave is expected to begin to steepen into a shock and the neutrino mass loss for each progenitor is shown in the legend (c.f. Tables 1 \& 2 of \citetalias{Fernandez18}).}
    \label{fig: mesa}
\end{figure*} 

The resolution to this apparent contradiction is that the relative mass loss and density power-law index are both position-dependent in a realistic stellar progenitor. To see this directly, the left panel of Figure \ref{fig: mesa} shows the radial density profiles of the $15\Msun$ RSG, $22\Msun$ YSG, and $25\Msun$ BSG progenitors investigated by \citetalias{Fernandez18} (see their Tables 1 \& 2 for the complete properties of these progenitors). Here mass is the zero age main sequence mass, both stars are solar metallically, and we evolved the progenitors to the onset of core-collapse using the stellar evolution code \textsc{mesa} version 6794 \citep{Paxton2011, Paxton2013, Paxton2015, Paxton2018, Paxton2019, Jermyn2023} with the publicly available inlists\footnote{\url{https://bitbucket.org/rafernan/bhsn_mesa_progenitors}} provided by \citetalias{Fernandez18} (see also \citealt{Fuller15}). At the base of the hydrogen envelope --- near $\sim3\times10^{11}$cm for each progenitor --- the RSG and YSG profiles are well-approximated by a $\rho \propto r^{-2.5}$ power-law (dashed curve), and this same power-law index is effectively sustained until the stellar surface for the YSG. The density profile of the RSG, on the other hand, flattens significantly near $\sim10^{12}$cm to $\rho \propto r^{-1.6}$ (dotted curve), while the BSG has a much smaller radius at core collapse than either the YSG or RSG and, thus, possesses a much more rapidly declining density profile overall. 
Consequently, the RSG hydrogen envelope contains a significant fraction of the total stellar mass, while it contains very little for the YSG and BSG. 

The corollary of this progenitor dependence is that, even though the absolute mass lost $\delta M$ and the total fractional mass loss $\delta M/M_{\star}$ are comparable for each of these stars, the {relative} mass loss at small radii (relative to the surface) in the stellar interior, $\delta M/M(r)$, is substantially larger for the RSG. The right panel of Figure \ref{fig: mesa} shows this directly for each of these stars, where the absolute mass loss (as presented in Table 2 of \citetalias{Fernandez18}) is given in the legend. Also indicated is the location within each progenitor where the wave is expected to steepen substantially (defined as $r_{\rm c}$ in \citetalias{Fernandez18} and equal to the radius in the star where the local freefall time is comparable to the neutrino diffusion time of 3 seconds) by the vertical dashed lines, which are $1.5$, $0.8$, \& $1.7\times10^9$cm for the RSG, YSG, and BSG, respectively. Consistent with expectations, the relative mass loss of the RSG progenitor is everywhere greater than that of the YSG and BSG, and larger by a factor of $\sim 2-3$ at the base of the hydrogen envelope (exterior to which $\delta M/M(r) \simeq const$.

Moreover, the density profile at radii where the shock initially forms can be much steeper than that appropriate to the outer hydrogen envelope, which is apparent from the left panel of Figure \ref{fig: mesa}. In particular, the average decline of the density between the wave steepening radius ($\sim 10^{9}$ cm) and the base of the hydrogen envelope ($\sim 3\times 10^{11}$ cm) is $\rho \propto r^{-3.8}$ (the dot-dashed curve in the left panel of Figure \ref{fig: mesa}). Not only is this substantially steeper than the $\propto r^{-2.5}$ or $\propto r^{-1.6}$ power-law falloff appropriate to the hydrogen envelope of the stars, it is steeper than the power-law decline at which \emph{any} weak-shock solution exists, being $\propto r^{-3.5}$ \citep{coughlin19}. The system may therefore be well in excess of the critical mass loss appropriate to the power-law index where the shock initially forms (potentially infinitely so), implying that it can rapidly approach the strong limit and accelerate according to the type-II similarity solution described in \citet{Waxman93}. 

This outcome -- that the system greatly exceeds the critical mass loss and the shock rapidly accelerates into the strong regime -- is characteristic of the RSG but not of the YSG: near $10^9$ cm both density profiles are well described by a $\rho \propto r^{-2.2}$ power-law, which (from the self-similar solutions and assuming at this radius $\gamma\simeq 4/3$) corresponds to a maximum mass loss of $\left(\delta M/M\right)_{\rm max}\simeq 0.081$, while the instantaneous relative mass losses for the RSG and YSG at the wave steepening radius (dashed lines) are $\delta M_{\rm RSG}/ M\left(r_{\rm c}\right)=0.10$ and $\delta M_{\rm YSG}/M\left(r_{\rm c}\right)=0.036$, respectively. The relative mass loss for the RSG is therefore super-critical ($\Delta M/M=0.019$), whereas it is sub-critical for the YSG, implying that the RSG (YSG) shock will (will not) strengthen with time. Similarly, once the shocks reach the base of the hydrogen envelope (prior to which the RSG shock speed will have increased, given the $\propto r^{-3.8}$ scaling of the density profile between $\sim 10^{10}$ cm and $\sim 3\times 10^{11}$ cm) where the RSG and YSG density profiles are $\propto r^{-2.5}$, the critical mass loss for $n=2.5$ and $\gamma=1.4$ is $\left(\delta M/M\right)_{\rm max}=0.029$. Evaluating the RSG and YSG relative mass losses near the base of the hydrogen envelope at a radius of $3\times10^{11}$cm then gives $\delta M_{\rm RSG}/M=0.050$ and $\delta M_{\rm YSG}/M=0.012$, which shows that the RSG is still well above the critical value while the YSG is not. Even though the RSG envelope flattens to $\rho \propto r^{-1.6}$ where one would expect the shock to no longer strengthen, which is consistent with Figure 4 in \citetalias{Fernandez18} (the Mach number in the second-row, left-most panel declines with radius as the shock moves from $\sim 10^{12}$ cm to $\sim few\times 10^{13}$ cm) and Figure 13 in \citealt{coughlin18}, the shock is sufficiently strong to unbind a substantial fraction of the envelope, while the YSG shock remains weak and ejects effectively no mass. 

The BSG is in between these two extreme cases: similarly to the YSG, the fractional mass loss near where the shock forms is below the critical value (the local density profile of the BSG near $\sim few\times 10^{9}$ cm is flatter than $\propto r^{-2.2}$, as can be seen from the left panel of Figure \ref{fig: mesa}), implying that it starts in the weak regime. However, the density profile of the BSG hydrogen envelope is much steeper than that of the RSG or YSG, and is better described by $\rho \propto r^{-3}$ near the base (we opted not to show this power-law in the left panel of Figure \ref{fig: mesa} to maintain readability). This corresponds to a maximum mass loss of $\sim 2-4\times 10^{-3}$ depending on the adiabatic index (see the right panel of Figure \ref{fig: dM max}), and thus the local relative mass loss is $\sim$ an order of magnitude above the critical value. Despite starting in the weak regime, it should nevertheless strengthen down the ever-steepening outer envelope and enter the strong regime, successfully ejecting some relatively small amount of mass -- simultaneously because the shock began weaker and because there is little mass contained in the envelope. These expectations are consistent with the results of \citetalias{Fernandez18}, who found that the BSG FSNe resulted in the ejection of $\sim 0.05 M_{\odot}$. 

The fact that RSGs produce stronger explosions (as compared to BSGs, YSGs, and WRs) has been noted and found by all numerical and analytical investigations of FSNe that considered a range of progenitors (e.g., \citetalias{Fernandez18}, \citealt{Coughlin18b, coughlin18, Ivanov21, Schneider23}). The origin of this relative strength has often been attributed to the longevity of the neutron star: a less compact progenitor (or stiffer equation of state) enables a more prolonged protoneutron star phase (i.e., prior to black hole formation) and a correspondingly greater mass lost to neutrinos. While we do not disagree with this conclusion -- a larger absolute mass loss should yield a stronger explosion -- our analysis suggests that it is the combination of the \emph{relative} mass loss, evaluated at the position of the rarefaction wave at the onset of the mass loss, and the instantaneous power-law index of the ambient medium at that same radius, that are most important for establishing the explosion energy. The legitimacy of this conclusion is perhaps most readily verified by comparing the outcomes of the RSG and BSG FSNe analyzed in \citetalias{Fernandez18}: the mass lost (as considered in \citetalias{Fernandez18}) to neutrinos was the same to within a factor of $\sim 1.25$, while the mass ejected (again as described in \citetalias{Fernandez18}) differed by $\sim$ two orders of magnitude, even though the mass able to be ejected -- defined to be the mass exterior to $r_{\rm c}$ -- was $M(r>r_{\rm c}) = 7.85\Msun$ for the RSG and $7.82\Msun$ for the BSG; this latter feature can be seen from the right panel of Figure \ref{fig: mesa}, as both curves decline substantially in moving from $r_{\rm c}$ to $\sim$ the base of the hydrogen envelope.

\section{Summary and Conclusions}
\label{sec: sum}
In this work we analyzed the stability of the self-similar solutions presented in \citet{coughlin23}, which describe the propagation of a weak shockwave that forms in response to the mass-energy radiated in neutrinos, $\delta M$ in a core-collapse supernova. For a given relative mass lost to neutrinos, $\delta M/M$ (where $M$ is the pre-mass-loss mass generating the gravitational field at the initial radius of the rarefaction wave), that is below a critical value that depends on the ambient power-law index $n$ (and, less strongly, on the thermodynamics that -- in our simplified treatment here -- is contained in the adiabatic index $\gamma$), there are two self-similar solutions that differ in their Mach number. In Section \ref{sec: local}, we showed that the lager Mach number (``strong shock'') solutions are unstable to radial perturbations, with those perturbations growing as weak power-laws with time, and that the smaller Mach number (``weak shock'') solutions are dynamically stable. We verified these analytical results with {\sc flash} simulations, which showed that a shock formed following the mass loss, the dynamics of which were well described by the self-similar solution alongside the perturbation from the lowest-order eigenvalue. 

We showed in Section \ref{sec: global} that scenarios with mass losses above the critical value can be described by perturbing the maximum mass loss solutions. In this case the perturbations are driven by the of excess mass loss --- i.e., the difference between the total mass loss and maximum mass loss --- and we demonstrated that they grow logarithmically with the shock position (and time). As there is no weak shock self-similar solution that exists above the critical mass loss, these solutions asymptotically strengthen over time and inevitably transition into the strong/Sedov-Taylor/energy-conserving regime. However, the logarithmic growth of the perturbations implies that, unless the excess mass loss is large, the shock takes extremely long to reach the strong limit, and instead it propagates at effectively a fixed Mach number (see Figure \ref{fig: R shock}).

In Section \ref{sec: implications} we discussed the implications of our analysis in the context of realistic stellar profiles and compared our findings to those in \citet{Fernandez18}. We suggest that the variety of FSN outcomes that arise from different progenitors --- either extended red supergiants (RSGs) with large mass ejection or compact supergiants such as blue supergiants (BSGs), yellow supergiants (YSGs), and Wolf-Rayets (WRs) with little to no ejecta --- can be explained not by the differences in absolute mass lost to neutrinos between the progenitors, but by the differences in the \textit{relative} mass loss at the radius in the star where the sound wave steepens into a shock, $\delta M/M(r_{\rm c})$ (where, adopting the notation from \citetalias{Fernandez18}, $r_{\rm c}$ denotes this radius in the star). In particular, Figure \ref{fig: mesa} illustrates that the RSG is well in excess of the critical mass loss at $r_{\rm c}$, implying that it will strengthen significantly prior to reaching the base of the hydrogen envelope, while the BSG is well below it, leading to a shock that remains weak until it reaches the rapidly declining density profile of the outer envelope. Therefore, even though both stars have nearly the same absolute mass loss and mass exterior to $r_{\rm c}$, the RSG creates a much more powerful explosion with significantly more ejecta. 

In a FSN, neutrino radiation during the proto-neutron star phase inevitably ceases once the compact object reaches the TOV limit and forms a black hole. We therefore focused on the solutions of \citetalias{coughlin23} that permit accretion onto the newly formed black hole at the origin. However, \citetalias{coughlin23} also showed that there are weak-shock solutions to Equations \eqref{ss cont}--\eqref{ss ent} (discussed in Appendix B of \citetalias{coughlin23}) that maintain zero mass flux at the origin. Such ``settling'' solutions (the velocity goes to zero at the origin) could manifest in, for example, the non-terminal eruptions of massive stars or the response of a circumbinary disk following the merger of a black hole binary (as discussed in \citetalias{coughlin23}). Similarly, these settling solutions could apply to a successful CCSN where a neutron star persists instead of a black hole once causal connectedness is established, but the weak shockwave will be overtaken by the strong neutron star bounce shock that ultimately drives the explosion. In any case, the most general solution must satisfy initial -- as well as boundary -- conditions and is therefore not self-similar, but can be described using the eigenmode analysis presented here when the initial state is not too far from the self-similar one, as is the case for the shock that forms from the steepening of the rarefaction wave.

In our analysis we assumed that the shock propagated indefinitely into an ambient medium with a density profile described by a single power-law, and neglected variations that naturally exist due to various nuclear burning regions within the star (see the left panel of Figure \ref{fig: mesa}). In a realistic FSN, once the shock reaches a region of the star characterized by a different power-law index $n$, we expect it to strengthen or weaken according to its Mach number and local relative mass loss. More specifically, if the Mach number is greater than the strong shock self-similar Mach number for the new power-law index $n$, then the shock will continue to strengthen into the strong regime in accordance with the results of our stability analysis. This transition between the weak and strong shock solutions (or vice versa) can likely be accounted for analytically using an interpolating method similar to the one used by \cite{Matzner99} for strong/energy-conserving self-similar solutions (see also \citealt{Linial21}). We also showed that there are regions within supergiant progenitors where the density declines more rapidly than $r^{-3.5}$ (again see the left panel of Figure \ref{fig: mesa}), such as the region exterior to the wave steepening radius $r_{\rm c}$ and near the surface of the star. In such steep density gradients, the \citetalias{coughlin23} solution does not exist, and the shock accelerates in accordance with the \cite{Waxman93} self-similar solution in the interior of the star or the planar solutions of \cite{Sakurai60} as it approaches the stellar surface. 

The initial strengthening of the rarefaction wave into a shock, the shock strengthening/weakening as it propagates over multiple decades in radius, and the eventual shock breakout have significant implications for the observable signatures of FSNe. We intend to investigate this process analytically and numerically using realistic stellar progenitors in future work.

\section*{Acknowledgments}
D.A.P.~acknowledges support from the National Science Foundation through the Graduate Research Fellowship Program under grant No.~CON05112. Any opinions, findings, and conclusions or recommendations expressed in this material are those of the authors and do not necessarily reflect the views of the National Science Foundation. E.R.C.~and D.A.P.~acknowledge support from NASA through the Astrophysics Theory Program, grant 80NSSC24K0897.

The software used in this work was developed in part by the DOE NNSA- and DOE Office of Science-
supported Flash Center for Computational Science at the University of Chicago and the University of
Rochester.
\software{FLASH \citep{fryxell00}, MESA \citep{Paxton2011, Paxton2013, Paxton2015, Paxton2018, Paxton2019, Jermyn2023}}

\bibliographystyle{aasjournal}
\bibliography{ref}

\appendix
\section{Equations}
\label{Equations}
\subsection{Local Stability Equations}
\label{sec: local eqs}
Inserting Equations \eqref{ps vel}--\eqref{ps p} into the fluid equations (Equations \eqref{cont}--\eqref{entropy}) yields the unperturbed (self-similar) equations, which are
\begin{align}
    \label{ss cont}
    &-ng_0-\xi\frac{\partial g_0}{\partial \xi}+\frac{1}{{\xi^2}}\frac{\partial }{\partial \xi}\left[{\xi}^2f_0 g_0\right]=0, \\
    &-\frac{1}{2}f_0+\left(f_0-\xi\right)\frac{\partial f_0}{\partial \xi}+\frac{1}{g_0}\frac{\partial h_0}{\partial \xi} =-\frac{9}{4}{\etash}^2\frac{1}{{\xi}^2}\left(1-\frac{\delta M}{M}\right), \\
   \label{ss ent}
   &n\gamma -n -1+\left(f_0-\xi\right)\frac{\partial }{\partial \xi}\ln\left(\frac{h_0}{{g_0}^{\gamma}}\right)=0,
\end{align}
and the first-order equations, which are given by
\begin{align}
    \label{pert cont}
    &-ng_1 +\frac{\partial g_1}{\partial \tau}-\xi\frac{\partial g_1}{\partial \xi}+\frac{1}{{\xi}^2}\frac{\partial}{\partial \xi}\left[\xi^2\left(f_0 g_1 + g_0 f_1\right)\right]=0, \\
    &-\frac{1}{2}f_1-\frac{1}{\etash}\left[\dot \eta_1 +\frac{2}{3}\ddot \eta_1\right]f_0+\frac{\partial f_1}{\partial \tau}-\xi\frac{\partial f_1}{\partial \xi}+\frac{\partial}{\partial \xi}\left[f_0 f_1\right]+\frac{1}{g_0}\left[\frac{\partial h_1}{\partial \xi}-\frac{g_1}{g_0}\frac{\partial h_0}{\partial \xi}\right]=-\frac{\etash}{\xi^2}\left[\frac{9}{2}\eta_1+3\dot \eta_1\right]\left(1-\frac{\delta M}{M}\right), \\
    \label{pert ent}
    &-\frac{2}{\etash}\left[\dot \eta_1+\frac{2}{3}\ddot \eta_1\right]+\left(f_0-\xi\right)\frac{\partial}{\partial \xi}\left[\frac{h_1}{h_0}-\gamma \frac{g_1}{g_0}\right]+f_1\frac{\partial }{\partial \xi}\ln\left(\frac{h_0}{{g_0}^{\gamma}}\right)+\frac{\partial }{\partial \tau}\left[\frac{h_1}{h_0}-\gamma \frac{g_1}{g_0}\right]=0. 
\end{align}
Taking the Laplace transform of of the first-order equations --- as defined by Equation \eqref{lt f} --- and re-defining the variables such that, for example, $\tilde{f}_1 \rightarrow \tilde{f}_1/\tilde{\eta}_1$ then gives
\begin{align}
    \label{lt pert cont}
    \left(\sigma-n\right)\tilde{g}_1-\xi \frac{\partial \tilde{g}_1}{\partial \xi}+\frac{1}{{\xi}^2}\frac{\partial}{\partial \xi}\left[\xi^2\left(f_0 \tilde{g}_1 + g_0 \tilde{f}_1\right)\right]&=0, \\
    -\frac{1}{2}\tilde{f}_1-\frac{1}{\etash}\left[\sigma +\frac{2}{3}{\sigma}^2\right]f_0+\sigma \tilde{f}_1-\xi \frac{\partial \tilde{f}_1}{\partial \xi}+\frac{\partial }{\partial \xi}\left[f_0 \tilde{f}_1\right]+\frac{1}{g_0}\left[\frac{\partial \tilde{h}_1}{\partial \xi}-\frac{\tilde{g}_1}{g_0}\frac{\partial h_0}{\partial \xi}\right]&=-\frac{\etash}{\xi^2}\left[\frac{9}{2}+3\sigma\right]\left(1-\frac{\delta M}{M}\right), \\
    -\frac{2}{\etash}\left[\sigma+\frac{2}{3}\sigma^2\right]+\left(f_0-\xi\right)\frac{\partial}{\partial \xi}\left[\frac{\tilde{h}_1}{h_0}-\gamma \frac{\tilde{g}_1}{g_0}\right]+\tilde{f}_1\frac{\partial }{\partial \xi}\ln\left(\frac{h_0}{{g_0}^{\gamma}}\right)+\sigma\left[\frac{\tilde{h}_1}{h_0}-\gamma \frac{\tilde{g}_1}{g_0}\right]&=0.
    \label{lt pert ent}
\end{align}

At the location of the shock, the continuity of mass, momentum, and energy fluxes across the shockfront require that the fluid velocity, density, and pressure evaluated at the shock satisfy
\begin{align}
    \label{vel cond}
    v\left(R_{\rm sh}\right)&=\frac{2}{\gamma+1}\left[1+\frac{\gamma-1}{2}\frac{v_{\rm e}}{V_{\rm sh}}-\frac{\gamma p_{\rm e}}{\rho_{\rm e}{V_{\rm sh}^2}\left(1-\frac{v_{\rm e}}{V_{\rm sh}}\right)}\right]V_{\rm sh}, \\
    \label{rho cond}
    \rho\left(R_{\rm sh}\right) &= \frac{\gamma+1}{\gamma-1}\left[1+\frac{2\gamma}{\gamma-1}\frac{p_{\rm e}}{\rho_{\rm e}{V_{\rm sh}^2}\left(1-\frac{v_{\rm e}}{V_{\rm sh}}\right)^2}\right]^{-1}\rho_{\rm e}, \\
    \label{p cond}
    p\left(R_{\rm sh}\right) &= \frac{2}{\gamma+1}\left[\left(1-\frac{v_{\rm e}}{V_{\rm sh}}\right)^2-\frac{\gamma-1}{2}\frac{p_{\rm e}}{\rho_{\rm e}{V_{\rm sh}}^2}\right]\rho_{\rm e}{V_{\rm sh}}^2.
\end{align}
Inserting Equations \eqref{amb vel}--\eqref{amb pres} into the above expressions and setting them equal to Equations \eqref{ps vel}--\eqref{ps p} evaluated at $\xi=1$ then provides the boundary conditions for the self-similar variables. The zeroth order boundary conditions are
\begin{align}
    \label{f0 bound}
    f_0\left(\xi = 1\right) &= \frac{2}{\gamma+1}+\frac{\gamma-1}{\gamma+1}\frac{3\etash}{2}f_{\rm e}-\frac{2\gamma}{\left(\gamma+1\right)\left(n+1\right)}\frac{9\etash^2}{4}\frac{h_{\rm e}}{g_{\rm e}}\left(1-\frac{3\etash}{2}f_{\rm e}\right)^{-1}, \\
    \label{g0 bound}
    g_0\left(\xi = 1\right) &= \frac{\gamma+1}{\gamma-1}\left(1+\frac{2\gamma}{\gamma-1}\frac{9\etash^2}{4}\frac{1}{n+1}\frac{h_{\rm e}}{g_{\rm e}}\times\left(1-\frac{3\etash}{2}f_{\rm e}\right)^{-2}\right)^{-1} g_{\rm e}, \\
    \label{h0 bound}
    h_0\left(\xi = 1\right) &= \left(\frac{2}{\gamma+1}\left(1-\frac{3\etash}{2}f_{\rm e}\right)^2-\frac{\gamma-1}{\gamma+1}\frac{9\etash^2}{4}\frac{1}{n+1}\frac{h_{\rm e}}{g_{\rm e}}\right)g_{\rm e},
\end{align}
while the Laplace transformed, first-order boundary conditions are
\begin{align}
    \label{f1 bc}
    \tilde{f}_1\left(\xi=1\right) = \frac{2}{1+\gamma }\left(\frac{1}{4} (-1+\gamma ) \left(3 f_{\rm e}+2 \sigma  f_{\rm e}+3 \etash f_{\rm e}'\right)+\frac{\gamma  h_{\rm e} \left(-\frac{9 \etash^2 \left(3 f_{\rm e}+2 \sigma  f_{\rm e}+3 \etash f_{\rm e}'\right)}{2 \left(-2+3 f_{\rm e} \etash\right){}^2}-\frac{3 \left(\frac{3}{2}+\sigma \right) \etash+\frac{9}{4} \etash^2 \left(-\frac{g_{\rm e}'}{g_{\rm e}}+\frac{h_{\rm e}'}{h_{\rm e}}\right)}{1-\frac{3}{2} f_{\rm e} \etash}\right)}{(1+n) g_{\rm e}}\right),
\end{align}
\begin{align}
\begin{split}
    \tilde{g}_1\left(\xi=1\right) &= \frac{\gamma +1}{\gamma -1}
        \left(\vphantom{\dfrac{g_e'}{g_e \left(\dfrac{18 \gamma \etash^2 h_e}
                                        {(\gamma -1)(n+1) g_e
                                         \left(3 \etash f_e-2\right){}^2}
                              +1\right)}}
        \frac{6 \gamma  \left(18 \etash^3 f_e' g_e h_e + 9 \etash^3 f_e g_e' h_e
              - 9 \etash^3 f_e g_e h_e' - 6 \etash^2 g_e' h_e
              + 6 \etash^2 g_e h_e' + 8 \sigma \eta_s g_e h_e
              + 12 \etash g_e h_e\right)}
             {(\gamma -1)(n+1) g_e{}^2 \left(3 \etash f_e-2\right){}^3
              \left(\dfrac{18 \gamma \etash^2 h_e}
                         {(\gamma -1)(n+1) g_e \left(3 \etash f_e-2\right){}^2}
                   +1\right){}^2}\right. \\
    &\qquad \left. +\frac{g_e'}{g_e \left(\dfrac{18 \gamma \etash^2 h_e}
                                        {(\gamma -1)(n+1) g_e
                                         \left(3 \etash f_e-2\right){}^2}
                              +1\right)}\right) g_e,
\end{split}
\end{align}
\begin{align}
\begin{split}
    \label{h1 bc}
    \tilde{h}_1\left(\xi=1\right) &= \frac{2}{1+\gamma}
        \left(\vphantom{
            \dfrac{(-1+\gamma) h_e \left(-3 \left(\frac{3}{2}+\sigma \right) \etash
                -\frac{9}{4} \etash^2 \left(-\frac{g_e'}{g_e}+\frac{h_e'}{h_e}\right)\right)}
                  {2 (1+n) g_e}}
        \left(1-\frac{3}{2} \etash f_e\right)
        \left(-3 f_e-2\sigma f_e-3\etash f_e'\right) \right.\\
    &\qquad \left.
        +\frac{\left(\left(1-\frac{3}{2}\etash f_e\right){}^2
            -\frac{9(-1+\gamma)\etash^2 h_e}{8(1+n)g_e}\right)g_e'}{g_e}
        +\frac{(-1+\gamma) h_e \left(-3\left(\frac{3}{2}+\sigma\right)\etash
            -\frac{9}{4}\etash^2\left(-\frac{g_e'}{g_e}+\frac{h_e'}{h_e}\right)\right)}
              {2(1+n)g_e}
        \right) g_e,
\end{split}
\end{align}
Here the ambient self-similar functions --- $f_{\rm e}$, $g_{\rm e}$, and $h_{\rm e}$ --- are evaluated at $\etash$ and primes denote derivatives with respect to $\eta$. Equations \eqref{ss cont}--\eqref{pert ent} can thus be integrated inward from the shock front ($\xi=1$) for a given mass loss $\delta M/M$, power-law index $n$, and adiabatic index $\gamma$ to solve for the corresponding eigenvalue $\sigma$. 

\subsection{Global Stability Equations}
\label{sec: global eqs}
Inserting Equations \eqref{ve sup crit}--\eqref{pe sup crit} into the fluid equations (with the momentum equation having the additional $\Delta M/M$ term to account for the total mass loss) yields the equations which describe perturbations to the ambient solutions evaluated at the maximum mass loss $\left(\delta M/M\right)_{\rm max}$, and are given by
\begin{align}
    \label{amb cont sup crit}
    &\frac{\partial g_{\rm e,1}}{\partial \eta}+\left(\frac{3}{2}-n\right)\left[f_{\rm e,0}g_{\rm e,1}+f_{\rm e,1}g_{\rm e,0}\right]-\frac{3}{2}\eta\frac{\partial}{\partial \eta}\left[f_{\rm e,0}g_{\rm e,1}+f_{\rm e,1}g_{\rm e,0}\right]=0, \\
    &\frac{\partial f_{\rm e,1}}{\partial \eta}-\frac{1}{2}\left[f_{\rm e,0}\left(f_{\rm e,1}+3\eta \frac{\partial f_{\rm e,1}}{\partial \eta}\right)+f_{\rm e,1}\left(f_{\rm e,0}+3\eta\frac{\partial f_{\rm e,0}}{\partial \eta}\right)\right]\nonumber \\
    &-\frac{1}{n+1}\frac{1}{g_{\rm e,0}}\left[\left(n+1\right)h_{\rm e,1}+\frac{3}{2}\eta \frac{\partial h_{\rm e,1}}{\partial\eta}-\frac{g_{\rm e,1}}{g_{\rm e,0}}\left\{\left(n+1\right)h_{\rm e,0}+\frac{3}{2}\eta\frac{\partial h_{\rm e,0}}{\partial\eta}\right\}\right]=1, \\
    &\frac{\partial}{\partial \eta}\left[\frac{h_{\rm e,1}}{h_{\rm e,0}}-\gamma\frac{g_{\rm e,1}}{g_{\rm e,0}}\right]+f_{\rm e,1}\left[n\gamma-n-1-\frac{3}{2}\eta\frac{\partial}{\partial \eta}\ln{\left(\frac{h_{\rm e,0}}{{g_{\rm e,0}^{\gamma}}}\right)}\right]-\frac{3}{2}\eta f_{\rm e,0}\frac{\partial}{\partial \eta}\left[\frac{h_{\rm e,1}}{h_{\rm e,0}}-\gamma\frac{g_{\rm e,1}}{g_{\rm e,0}}\right]=0, \label{amb entropy sup crit}
\end{align}
which, to retain hydrostatic balance at $t=0$, have the initial conditions $f_{\rm e,1}\left(0\right)=g_{\rm e,1}\left(0\right)=h_{\rm e,1}\left(0\right)=0$. 

The zeroth order equations are again given by Equations \eqref{ss cont}--\eqref{ss ent}, while the Laplace transformed, first-order dimensionless fluid equations are
\begin{align}
    \label{cont sup crit}
    &\left(\sigma-n\right){\tilde{g}^{\dagger}_1}-\xi \frac{\partial {\tilde{g}^{\dagger}_1}}{\partial \xi}+\frac{1}{{\xi}^2}\frac{\partial}{\partial \xi}\left[\xi^2\left(f_0 {\tilde{g}^{\dagger}_1} + g_0 {\tilde{f}^{\dagger}_1}\right)\right]=0, \\
    &-\frac{1}{2}{\tilde{f}^{\dagger}_1}-\frac{1}{\etash}\left[\sigma +\frac{2}{3}{\sigma}^2\right]f_0+\sigma {\tilde{f}^{\dagger}_1}-\xi \frac{\partial {\tilde{f}^{\dagger}_1}}{\partial \xi}+\frac{\partial }{\partial \xi}\left[f_0 {\tilde{f}^{\dagger}_1}\right]+\frac{1}{g_0}\left[\frac{\partial {\tilde{h}^{\dagger}_1}}{\partial \xi}-\frac{{\tilde{g}^{\dagger}_1}}{g_0}\frac{\partial h_0}{\partial \xi}\right] \nonumber \\
    &=-\frac{1}{\xi^2}\left[\etash\left(\frac{9}{2}+3\sigma\right)\left(1-\frac{\delta M}{M}\right)-\frac{9}{4}\frac{\etash^2}{{\tilde{\eta}^{\star}_1}}\right], \\
    \label{ent sup crit}
    &-\frac{2}{\etash}\left[\sigma+\frac{2}{3}\sigma^2\right]+\left(f_0-\xi\right)\frac{\partial}{\partial \xi}\left[\frac{{\tilde{h}^{\dagger}_1}}{h_0}-\gamma \frac{{\tilde{g}^{\dagger}_1}}{g_0}\right]+{\tilde{f}^{\dagger}_1}\frac{\partial }{\partial \xi}\ln\left(\frac{h_0}{{g_0}^{\gamma}}\right)+\sigma\left[\frac{{\tilde{h}^{\dagger}_1}}{h_0}-\gamma \frac{{\tilde{g}^{\dagger}_1}}{g_0}\right]=0, 
\end{align}
where here we re-defined the dimensionless functions such that, for example, 
\begin{equation}
    {\tilde{f}^{\dagger}_1}\equiv \frac{\tilde{f_1}}{\tilde{\eta}_1},
\end{equation}
and 
\begin{equation}
    \tilde{\eta}_1^{\star}=\sigma \tilde{\eta}_1.
\end{equation}
At the shock front, the first-order functions satisfy
\begin{align}
\begin{split}
    {\tilde{f}^{\dagger}_1}\left(\xi=1\right)&=\frac{1}{2(1+n)(1+\gamma)\left(2 - 3\eta_{\rm sh} f_{\rm e,0}\right)^2 g_{\rm e,0}^2}
        \left[\vphantom{\dfrac{f_{\rm e,1}}{\tilde{\eta}^{\star}_1}}
        9(1+n)(\gamma-1)\eta_{\rm sh}^2(3+2\sigma) f_{\rm e,0}^3 g_{\rm e,0}^2 \right.\\
    &\qquad \left.
        + 3(1+n)(\gamma-1)\eta_{\rm sh} f_{\rm e,0}^2 g_{\rm e,0}^2
        \left(-4(3+2\sigma) + 9\eta_{\rm sh}^2
        \left(\frac{f_{\rm e,1}}{\tilde{\eta}^{\star}_1} + f_{\rm e,0}'\right)\right) \right.\\
    &\qquad \left.
        + 2 f_{\rm e,0}
        \left(\vphantom{\dfrac{h_{\rm e,1}}{\tilde{\eta}^{\star}_1}}
        2(1+n)(\gamma-1) g_{\rm e,0}^2
        \left(3+2\sigma - 9\eta_{\rm sh}^2
        \left(\frac{f_{\rm e,1}}{\tilde{\eta}^{\star}_1} + f_{\rm e,0}'\right)\right)
        \right.\right.\\
    &\qquad\qquad \left.\left.
        - 27\gamma\eta_{\rm sh}^3 h_{\rm e,0}
        \left(\frac{g_{\rm e,1}}{\tilde{\eta}^{\star}_1} + g_{\rm e,0}'\right)
        + 9\gamma\eta_{\rm sh}^2 g_{\rm e,0}
        \left((3+2\sigma) h_{\rm e,0}
        + 3\eta_{\rm sh}\left(\frac{h_{\rm e,1}}{\tilde{\eta}^{\star}_1} + h_{\rm e,0}'\right)\right)
        \right)\right.\\
    &\qquad \left.
        - 6\eta_{\rm sh}
        \left(\vphantom{\dfrac{f_{\rm e,1}}{\tilde{\eta}^{\star}_1}}
        \frac{f_{\rm e,1} g_{\rm e,0}
            \left(-2(1+n)(\gamma-1) g_{\rm e,0} + 9\gamma\eta_{\rm sh}^2 h_{\rm e,0}\right)}
            {\tilde{\eta}^{\star}_1}
        - 2(1+n)(\gamma-1) g_{\rm e,0}^2 f_{\rm e,0}'
        \right.\right.\\
    &\qquad\qquad \left.\left.
        - 6\gamma\eta_{\rm sh} h_{\rm e,0}
        \left(\frac{g_{\rm e,1}}{\tilde{\eta}^{\star}_1} + g_{\rm e,0}'\right)
        + \gamma g_{\rm e,0}
        \left(h_{\rm e,0}\left(12+8\sigma+9\eta_{\rm sh}^2 f_{\rm e,0}'\right)
        + 6\eta_{\rm sh}\left(\frac{h_{\rm e,1}}{\tilde{\eta}^{\star}_1} + h_{\rm e,0}'\right)\right)
        \right)\right],
\end{split}
\end{align}
\begin{align}
\begin{split}
    {\tilde{g}^{\dagger}_1}\left(\xi=1\right)&=\frac{(1+n)(1+\gamma)(3\eta_{\rm sh} f_{\rm e,0}-2) g_{\rm e,0}}
          {\left((1+n)(\gamma-1)(2-3\eta_{\rm sh} f_{\rm e,0})^2 g_{\rm e,0}
           + 18\gamma\eta_{\rm sh}^2 h_{\rm e,0}\right)^2}
     \left[\vphantom{\dfrac{(1+n)(\gamma-1)(3\eta_{\rm sh} f_{\rm e,0}-2)^3 g_{\rm e,1}}
                           {\tilde{\eta}^{\star}_1}}
     36\gamma\eta_{\rm sh}^2(3\eta_{\rm sh} f_{\rm e,0}-2) h_{\rm e,0}
     \left(\frac{g_{\rm e,1}}{\tilde{\eta}^{\star}_1} + g_{\rm e,0}'\right) \right.\\
    &\qquad \left.
     + g_{\rm e,0}
     \left(\vphantom{\dfrac{18\gamma\eta_{\rm sh}^2 h_{\rm e,1}}{\tilde{\eta}^{\star}_1}}
     \frac{(1+n)(\gamma-1)(3\eta_{\rm sh} f_{\rm e,0}-2)^3 g_{\rm e,1}}{\tilde{\eta}^{\star}_1}
     + 12\gamma\eta_{\rm sh} h_{\rm e,0}
     \left(6+4\sigma + 9\eta_{\rm sh}^2
     \left(\frac{f_{\rm e,1}}{\tilde{\eta}^{\star}_1} + f_{\rm e,0}'\right)\right)
     \right.\right.\\
    &\qquad\qquad \left.\left.
     + (3\eta_{\rm sh} f_{\rm e,0}-2)
     \left(-\frac{18\gamma\eta_{\rm sh}^2 h_{\rm e,1}}{\tilde{\eta}^{\star}_1}
     + (1+n)(\gamma-1)(2-3\eta_{\rm sh} f_{\rm e,0})^2 g_{\rm e,0}'
     - 18\gamma\eta_{\rm sh}^2 h_{\rm e,0}'\right)
     \right)\right], \rm and
\end{split}
\end{align}
\begin{align}
\begin{split}
    {\tilde{h}^{\dagger}_1}\left(\xi=1\right)&=\frac{1}{4(1+n)(1+\gamma)}
    \left[\vphantom{\dfrac{g_{\rm e,1}}{\tilde{\eta}^{\star}_1}}
    -\frac{24(1+n)\eta_{\rm sh} f_{\rm e,1} g_{\rm e,0}}{\tilde{\eta}^{\star}_1}
    + \frac{8(1+n) g_{\rm e,1}}{\tilde{\eta}^{\star}_1}
    - 6(\gamma-1)\eta_{\rm sh}(3+2\sigma) h_{\rm e,0}
    + \frac{9(1-\gamma)\eta_{\rm sh}^2 h_{\rm e,1}}{\tilde{\eta}^{\star}_1}
    \right.\\
    &\qquad\left.
    - 24(1+n)\eta_{\rm sh} g_{\rm e,0} f_{\rm e,0}'
    + 8(1+n) g_{\rm e,0}'
    + 6(1+n)\eta_{\rm sh} f_{\rm e,0}^2
    \left((6+4\sigma) g_{\rm e,0}
    + 3\eta_{\rm sh}\left(\frac{g_{\rm e,1}}{\tilde{\eta}^{\star}_1} + g_{\rm e,0}'\right)\right)
    \right.\\
    &\qquad\left.
    - 4(1+n) f_{\rm e,0}
    \left(\vphantom{\dfrac{g_{\rm e,1}}{\tilde{\eta}^{\star}_1}}
    g_{\rm e,0}\left(6+4\sigma
    - 9\eta_{\rm sh}^2\left(\frac{f_{\rm e,1}}{\tilde{\eta}^{\star}_1}
    + f_{\rm e,0}'\right)\right)
    + 6\eta_{\rm sh}\left(\frac{g_{\rm e,1}}{\tilde{\eta}^{\star}_1} + g_{\rm e,0}'\right)
    \right)
    - 9(\gamma-1)\eta_{\rm sh}^2 h_{\rm e,0}'
    \right].
\end{split}
\end{align}
Here primes denote derivatives of the ambient solutions denote derivatives with respect to $\eta$, and all ambient solutions are evaluated at $\etash$.

Upon specifying a value of for the eigenvalue $\sigma$ and perturbation ${\tilde{\eta}^{\star}_1}$, Equations \eqref{cont sup crit}--\eqref{ent sup crit} can therefore be integrated numerically with the boundary conditions to solve for the dimensionless perturbed functions. However, as noted in Section \ref{sec: local}, the critical mass loss solutions have a specific eigenvalue of $\sigma=0$ and, at this eigenvalue, the value of ${\tilde{\eta}^{\star}_1}$ diverges. The perturbations can thus be reconstructed by writing ${\tilde{\eta}^{\star}_1}$ as a sum over poles in the complex plane (see \citealt{Coughlin20}, for example). Specifically, we can write
\begin{equation}
   \label{poles}
   {\tilde{\eta}^{\star}_1}\equiv \sigma \tilde{\eta}_1 = \sum_{i}\frac{c_i}{\sigma -\sigma_i}.
\end{equation}
As the critical mass loss solution is characterized by a single eigenvalue of $\sigma=0$, we can re-arrange and differentiate Equation \eqref{poles} to define
\begin{equation}
   \label{c1}
   c_1=\left(\frac{\partial}{\partial \sigma}\left[\frac{1}{\tilde{\eta}^{\star}_1}\right]\bigg\lvert_{\sigma=0}\right)^{-1}.
\end{equation}
Therefore, the coefficient $c_1$ can be calculated by evaluating $({\tilde{\eta}^{\star}_1})^{-1}$ in the immediate vicinity of the $\sigma=0$. The temporal evolution of the perturbations can then be found by taking the inverse Laplace Transform of our functions, which are given by, for example,
\begin{equation}
   \label{inverse lt}
   \eta_1 = \frac{1}{2\pi i}\oint {\tilde{\eta}^{\star}_1}\left(\sigma\right)e^{\sigma \tau}d\sigma =\frac{1}{2\pi i}\oint \frac{c_1}{\sigma^2}e^{\sigma \tau}d\sigma,
\end{equation}
which has a second order pole at $\sigma=0$. Using the Residue Theorem we can evaluate the contour integral in Equation \eqref{inverse lt}, which gives 
\begin{align} 
   \label{eta tau}
   \eta_1&=\frac{1}{2\pi i}\times 2\pi i \times \rm Res \nonumber \\
   &= \lim_{\sigma \to 0}\frac{d}{d\sigma}\left[\sigma^2 \frac{c_1}{\sigma^2}e^{\sigma\tau}\right]=c_1\tau.
\end{align}
Perturbations to the maximum mass loss self-similar solutions therefore grow logarithmically with shock position (and time).

\section{Non-Radial Perturbations}
\label{sec: Angular}
The perturbation analysis presented in Sections \ref{sec: local} \& \ref{sec: global} was restricted to radial (i.e., spherically symmetric) perturbations, in which case we showed that the strong shock solutions are unstable while the weak solutions are stable. However, it is plausible that the shock is unstable to angular perturbations induced by, for example, convective motions in the hydrogen envelopes of massive stars. Account for angular perturbations to the shock propagation can be accomplished by writing corrections to the fluid quantities (ambient and post-shock) as spherical harmonics and accounting for the non-radial velocity components (e.g., \cite{Ryu87}). For example, the corrections imposed to the self-similar shock position and velocity for the sub-critical mass loss solutions (Section \ref{sec: local}) are given by
\begin{align}
    \frac{\sqrt{GM}t}{{R_{\rm sh}}^{3/2}}&=\etash+Y_\ell^m \eta_1\left(\tau\right), \\
    \frac{dR_{\rm sh}}{dt}&\equiv V_{\rm sh} = \sqrt{\frac{GM}{R_{\rm sh}}}\left[Y_\ell^m\dot{\eta}_1+\frac{3}{2}\left(\etash + Y_\ell^m\eta_1 \right)\right]^{-1},
\end{align}
and Equations \eqref{ps vel}-\eqref{ps p} become
\begin{align}
    \label{ps vel ang}
    v &= V_{\rm sh}\left[f_0\left(\xi\right)+Y_\ell^m f_1\left(\xi,\tau\right)\right], \\
    \label{ps rho ang}
    \rho &= \rho_{\rm i}{\left(\frac{R_{\rm sh}}{r_{\rm i}}\right)}^{-n}\left[g_0\left(\xi\right)+Y_\ell^m g_1\left(\xi,\tau\right)\right], \\
    \label{ps p ang}
    p &= \rho_{\rm i}{\left(\frac{R_{\rm sh}}{r_{\rm i}}\right)}^{-n}{V_{\rm sh}}^2\left[h_0\left(\xi\right)+Y_\ell^m h_1\left(\xi,\tau\right)\right].
\end{align}
Here $\tau$ retains the same definition as the log of the shock position given by Equation \eqref{tau}, and it is therefore an angular dependent quantity. We additionally define the angular components of the fluid velocity as
\begin{align}
    \label{v theta}
    v_{\theta}&=V_{\rm sh} f_{\perp}\frac{\partial Y_\ell^m}{\partial \theta}, \\
    \label{v phi}
    v_{\phi}&=V_{\rm sh}f_{\perp}\frac{1}{\sin\theta}\frac{\partial Y_\ell^m}{\partial \phi}.
\end{align}
We define perturbations to the ambient solution in the same way (i.e., all first-order quantities are $\propto Y_\ell^m$).

The fluid equations in spherical coordinates and keeping only linear terms in $\theta$ and $\phi$ are
\begin{align}
    \frac{\partial \rho}{\partial t}&+\frac{1}{r^2}\frac{\partial}{ \partial r}\left[r^2 \rho v_{\rm r}\right]+\frac{1}{r\sin{\theta}}\frac{\partial}{\partial \theta}\left[\rho v_{\theta}\sin{\theta}\right]+\frac{1}{r\sin{\theta}}\left[\rho v_{\phi}\right]=0, \\
    \frac{\partial v_{\rm r}}{\partial t}&+v_{\rm r}\frac{\partial v_{\rm r}}{\partial r}+\frac{1}{\rho}\frac{\partial p}{\partial r}=-\frac{GM}{r^2}, \\
    \frac{\partial v_{\theta}}{\partial t}&+v_{\rm r}\frac{\partial v_{\theta}}{\partial r}+\frac{v_{\rm r}v_{\theta}}{r}+\frac{1}{r}\frac{1}{\rho}\frac{\partial p}{\partial \theta}=0, \\
    \frac{\partial v_{\phi}}{\partial t}&+v_{\rm r}\frac{\partial v_{\phi}}{\partial r}+\frac{v_{\rm r}v_{\phi}}{r}+\frac{1}{r\sin{\theta}}\frac{1}{\rho}\frac{\partial p}{\partial \phi}=0, \\
    \frac{\partial s}{\partial t}&+v_{\rm r}\frac{\partial s}{\partial r}=0.
\end{align}

The zeroth-order equations remain the same as Equations \eqref{ss cont}-\eqref{ss ent}, however the first-order equations obtain a correction from the non-radial components of the fluid velocity, and are given by
\begin{align}
    &-ng_1 +\frac{\partial g_1}{\partial \tau}-\xi\frac{\partial g_1}{\partial \xi}+\frac{1}{{\xi}^2}\frac{\partial}{\partial \xi}\left[\xi^2\left(f_0 g_1 + g_0 f_1\right)\right]-\frac{l\left(l+1\right)}{\xi}g_0 f_{\perp}=0, \\
    &-\frac{1}{2}f_1-\frac{1}{\etash}\left[\dot \eta_1 +\frac{2}{3}\ddot \eta_1\right]f_0+\frac{\partial f_1}{\partial \tau}-\xi\frac{\partial f_1}{\partial \xi}+\frac{\partial}{\partial \xi}\left[f_0 f_1\right]+\frac{1}{g_0}\left[\frac{\partial h_1}{\partial \xi}-\frac{g_1}{g_0}\frac{\partial h_0}{\partial \xi}\right]=-\frac{\etash}{\xi^2}\left[\frac{9}{2}\eta_1+3\dot \eta_1\right]\left(1-\frac{\delta M}{M}\right), \\
    &-\frac{1}{2}f_{\perp}+\frac{\partial f_{\perp}}{\partial \tau}+\left(f_0-\xi\right)\frac{\partial f_{\perp}}{\partial \xi}+\frac{f_0 f_{\perp}}{\xi}+\frac{1}{g_0 \xi}\left[h_1+\frac{2}{3\etash}\left\{nh_0\eta_1-2h_0\left(\frac{3}{2}\eta_1+\dot \eta_1\right)+\eta_1 \xi \frac{\partial h_0}{\partial \xi}\right\}\right]=0, \\
    &-\frac{2}{\etash}\left[\dot \eta_1+\frac{2}{3}\ddot \eta_1\right]+\left(f_0-\xi\right)\frac{\partial}{\partial \xi}\left[\frac{h_1}{h_0}-\gamma \frac{g_1}{g_0}\right]+f_1\frac{\partial }{\partial \xi}\ln\left(\frac{h_0}{{g_0}^{\gamma}}\right)+\frac{\partial }{\partial \tau}\left[\frac{h_1}{h_0}-\gamma \frac{g_1}{g_0}\right]=0.
\end{align}
Here our system of equations is reduced by one equation because the $\theta$ and $\phi$ momentum equations are equivalent in dimensionless form, and we used the fact that 
\begin{equation}
    \frac{1}{\sin\theta}\frac{\partial}{\partial \theta}\left[\sin\theta\frac{\partial Y_\ell^m}{\partial \theta}\right]+\frac{1}{\sin^2\theta}\frac{\partial^2Y_\ell^m}{\partial \phi^2}=-l\left(l+1\right)Y_\ell^m
\end{equation}
to eliminate the angular dependence in the continuity equation. This set of equations can now be Laplace transformed using the same procedure in the previous sections, resulting in a system of ordinary differential equations that can be numerically integrated with the appropriate boundary conditions to solve for the eigenvalue $\sigma$, which again is is constrained by the regularity of the fluid variables through the sonic point. 

The boundary conditions for the radial velocity, density, and pressure at the shock front are given by Equations \eqref{f1 bc}--\eqref{h1 bc}, and the additional boundary condition for the angular velocity components is given by 
\begin{equation}
    v_{\theta}\left(R_{\rm sh}\right)=-\frac{2}{\gamma+1}\left(1-\frac{v_{\rm a}}{V_{\rm sh}}\right)\frac{V_{\rm sh}}{R_{\rm sh}}\frac{\partial R_{\rm sh}}{\partial \theta},
\end{equation}
which in terms of our self-similar variables is 
\begin{equation}
    \tilde{f}_{\perp}\left(\xi=1\right)= \frac{2}{\gamma+1}\left(\frac{2}{3\etash}-f_{\rm e}\left(\etash\right)\right).
\end{equation}
One could similarly use the boundary condition for the $\phi$ velocity component, however, for the same reason as above, the resulting boundary condition for $f_{\perp}$ would be the same as they only differ by a factor of $\sin\theta^{-1}$ when $m=0$.

\begin{figure*}
    \includegraphics[width=0.496\textwidth]{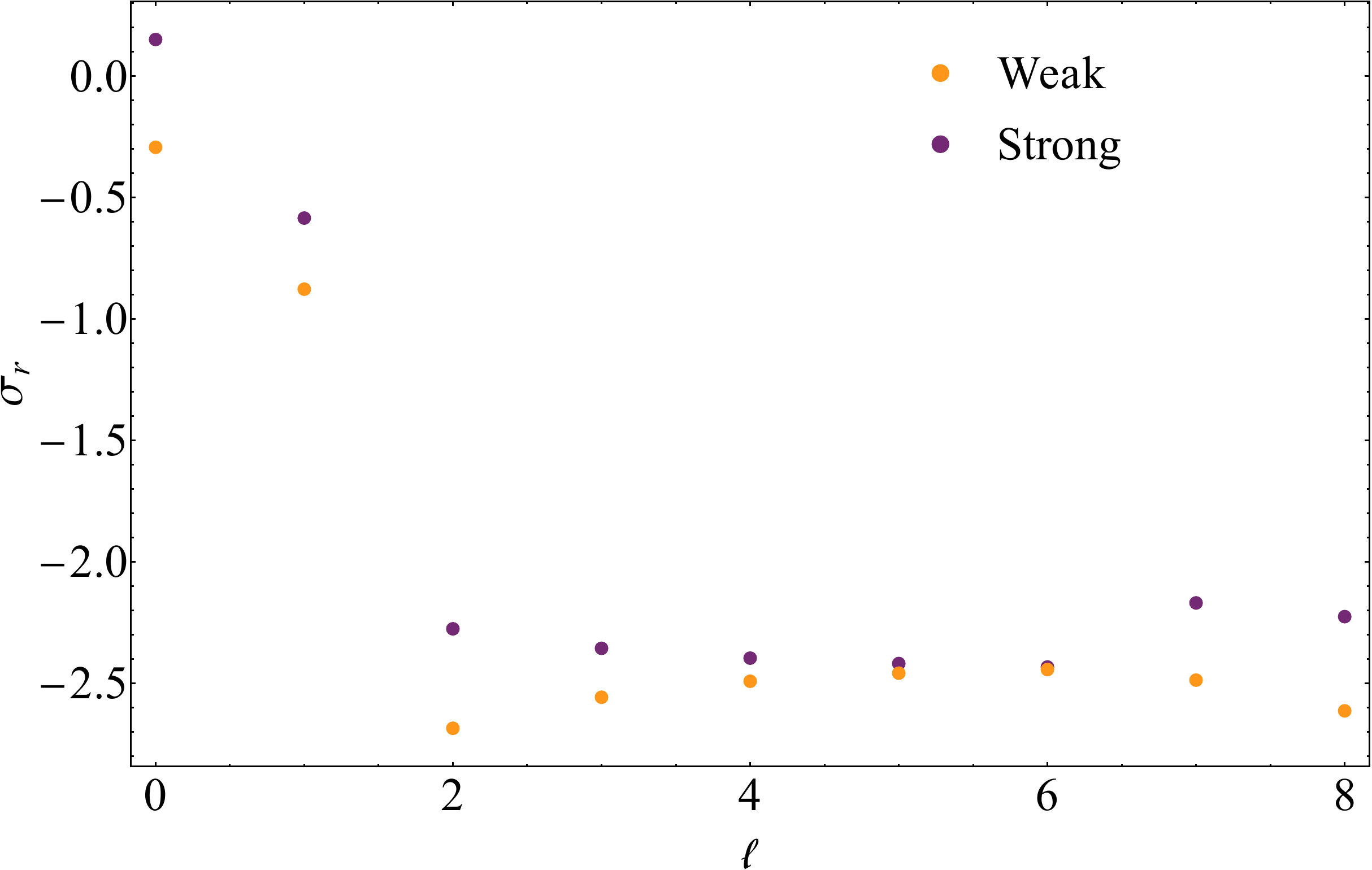}
    \includegraphics[width=0.498\textwidth]{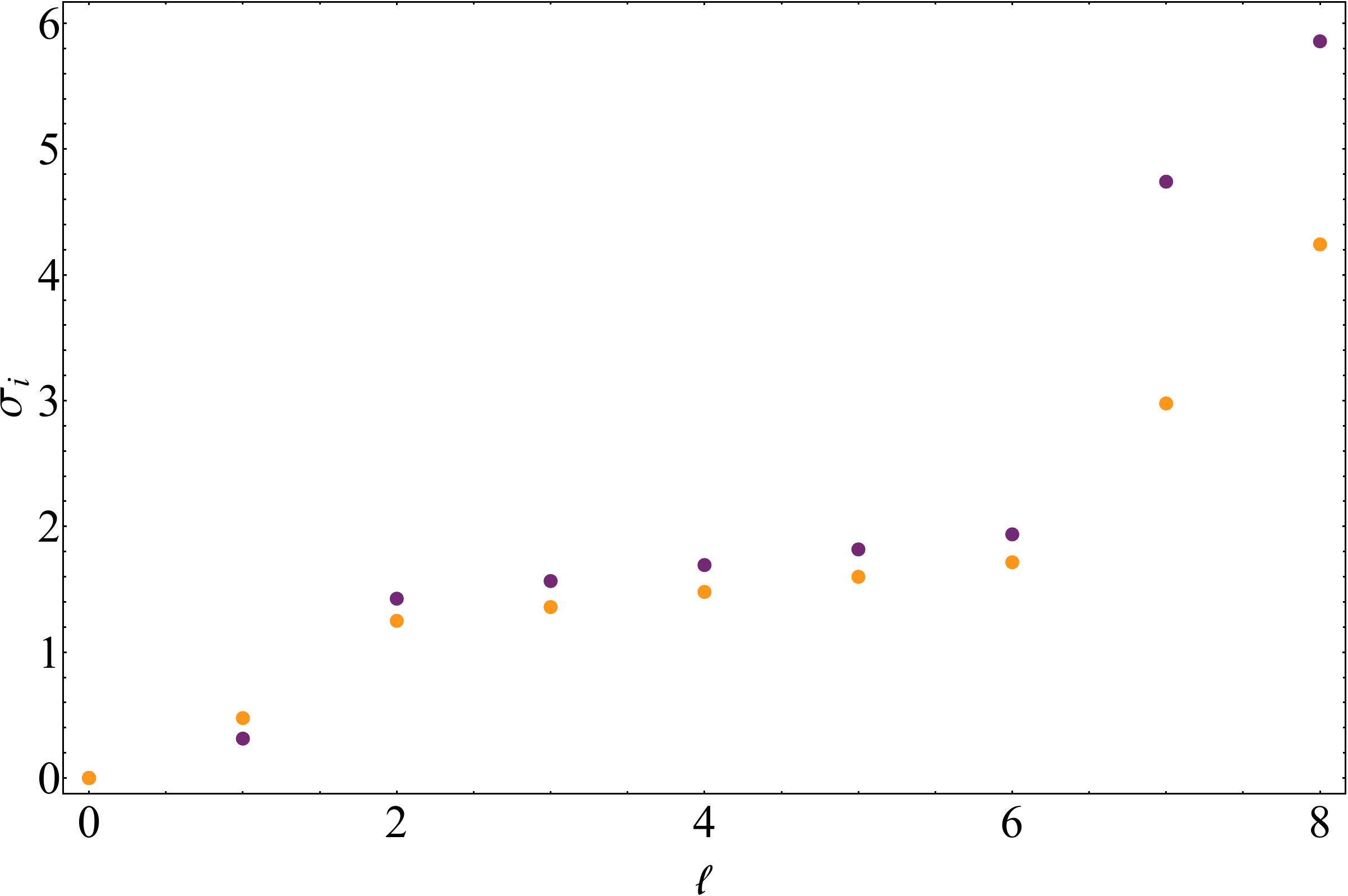}
    \caption{Left: The real part of the lowest-order (in magnitude) eigenvalue as a function of spherical harmonic $\ell$ for the weak and strong shock solutions for $n=2.5$, $\gamma=1.4$, and $\delta M/M=0.01$. It can be seen that only the $\ell=0$ mode is positive (unstable) for the strong shock solution, while the weak shock solution is stable for all $\ell$. Right: The imaginary part of the lowest-order eigenvalue as a function of $\ell$.}
    \label{fig: ang sigmas}
\end{figure*}
The left panel of Figure \ref{fig: ang sigmas} shows the real part of the lowest-order eigenvalue for the weak and strong shock solutions as a function of spherical harmonic $\ell$ for $n=2.5$, $\gamma=1.4$, and $\delta M/M=0.01$. As the $\ell=0$ solution is by definition the same eigenvalue from our purely radial perturbation analysis, we naturally recover the weak and strong shock eigenvalues presented in Section \ref{sec: local}. However, this figure shows that for $\ell>0$, both the weak and strong shock eigenvalue are negative, and are therefore stable. We have additionally verified this for other variations of $n$, $\gamma$, and $\delta M/M$. The right panel of the same figure shows the imaginary component of the lowest-order eigenvalues again as a function of $\ell$, where we have defined the eigenfrequency to be $\sigma=\sigma_{\rm r}+i\sigma_{\rm i}$. Modes with $\sigma_{\rm i}\neq0$ therefore induce oscillations to the shock position, velocity, and fluid quantities, while the real part controls the asymptotic growth of the perturbation. As noted in Section \ref{sec: local}, this shows that the lowest-order $\ell=0$ modes are purely real.

\end{document}